\documentclass[preprint,nofootinbib,amsmath,amssymb]{revtex4-1}
\usepackage{amsmath}
\usepackage{amssymb}
\usepackage{gensymb}
\usepackage{xargs}
\usepackage{dsfont}
\usepackage{color}
\usepackage{amsfonts}
\usepackage[normalem]{ulem}

\usepackage{bm}
\usepackage{graphicx}
\usepackage{hyperref}

\usepackage{epstopdf, epsfig}
\usepackage[utf8]{inputenc}
\usepackage{upgreek}
\usepackage{soul}
\newcommand{\dif}{\textup d}

\allowdisplaybreaks[1]

\begin{document}

\title{Attosecond betatron radiation pulse train}

\author{Vojtěch Horný}
\email{vojtech.horny@chalmers.se}
\affiliation{Department of Physics, Chalmers University of Technology, 412 96 Gothenburg, Sweden}
\affiliation{Institute of Plasma Physics, Czech Academy of Sciences, Za Slovankou 1782/3, 182 00 Praha 8, Czech Republic}
\author{Miroslav~Krůs}
\affiliation{Institute of Plasma Physics, Czech Academy of Sciences, Za Slovankou 1782/3, 182 00 Praha 8, Czech Republic}
\author{Wenchao Yan}
\affiliation{Institute of Physics, Czech Academy of Sciences, ELI BEAMLINES, Na Slovance 1999/2, 182 21 Praha 8, Czech Republic}
\affiliation{Key Laboratory for Laser Plasmas (MOE), School of Physics and Astronomy, Shanghai Jiao Tong University, Shanghai 200240, China}
\author{Tünde Fülöp}
\affiliation{Department of Physics, Chalmers University of Technology, 412 96 Gothenburg, Sweden}

\date{\today}

\begin{abstract}
High-intensity X-ray sources are essential diagnostic tools for science, technology and medicine. Such X-ray sources can be produced in laser-plasma accelerators, where electrons emit short-wavelength radiation due to their betatron oscillations in the  plasma wake of a laser pulse. Contemporary available betatron radiation X-ray sources can deliver a collimated X-ray pulse of duration on the order of several femtoseconds from a source size of the order of several micrometres. In this paper we demonstrate, through  particle-in-cell simulations, that the temporal resolution of such a source can be enhanced by an order of magnitude by a spatial modulation of the emitting relativistic electron bunch. The modulation is achieved by the interaction of the that electron bunch with a co-propagating laser beam which results in the generation of a train of equidistant sub-femtosecond X-ray pulses. The distance between the single pulses of a train is tuned by the wavelength of the modulation laser pulse. The modelled experimental setup is achievable with current technologies. Potential applications include stroboscopic sampling of ultrafast fundamental processes.
\end{abstract}

\maketitle


\section*{Introduction}
Sub-femtosecond high brightness X-ray pulses are in high demand by research communities in the fields of biology, material science or
femtochemistry \cite{martin2004femtochemistry}, as well as by industry
and medicine\cite{gotzfried2018research}. Such pulses can be used as a
diagnostic tool to resolve the structure and dynamics of dense matter,
proteins, and study fundamental physical phenomena such as chemical
reactions, lattice vibrations or phase transitions. Currently, high
brightness X-ray sources are produced by large scale facilities based
on radiation emission by relativistic electron bunches,
e.g.~synchrotron light sources \cite{bilderback2005review} and X-ray
free electron lasers \cite{mcneil2010x}. This limits their general
availability for many of the potential users. Here, we propose a new method to produce a train of equidistant sub-femtosecond X-ray pulses with a currently available laser systems.

Acceleration of electron bunches by the plasma wakefield driven by
laser \cite{esarey2009physics, gonsalves2019petawatt}, electron
\cite{blumenfeld2007energy}, or proton \cite{adli2018acceleration}
beams provides a promising alternative to the aforementioned concepts. The major advantage of plasma based accelerators is their ability to
sustain acceleration gradients of the order of hundreds of GeV/m,
which is approximately three orders of magnitude higher than is attainable with standard radiofrequency accelerators. Thus, the
electrons can be accelerated to energies of the order of hundreds of
MeV in a few millimeters. During the acceleration process, the
electron bunch undergoes transverse betatron oscillations due to the
presence of the transverse electric field. As a result, betatron radiation
\cite{kiselev2004x, rousse2004production, schnell2015characterization}
with a synchrotron-like \cite{fourmaux2011demonstration} spectrum,
typically in the X-ray range, is emitted.

The betatron radiation characteristics depend on the electron Lorentz
factor $\gamma$, plasma electron density $n_e$, betatron oscillation
amplitude $r_\beta$, and number of oscillation periods $N_0$. The
radiation spectrum is characterized by a critical energy, close to the
peak of the synchrotron spectrum, given in practical units
$\hbar \omega_c \textnormal{ [eV]} = 5.24\times 10^{-21} \gamma^2 n_e
\textnormal{ [cm\textsuperscript{-3}]} r_\beta \textnormal{
  [$\upmu$m]}$. The average photon number with energy $\hbar \omega_c$ emitted by an electron is $N_X = 5.6\times 10^{-3} N_0K$,
where
$K= 1.33 \times 10^{-10} \gamma^{1/2} n_e^{1/2} \textnormal{
  [cm\textsuperscript{-3}]} r_\beta \textnormal{ [$\upmu$m]}$ is the
strength parameter \cite{rousse2007scaling,
  corde2013femtosecond}. Several applications of such betatron sources
have been demonstrated, e.g.~diagnosing biological
samples\cite{Cole2015} and probing extreme states of
matter\cite{Albert_2016}, but others would require higher photon
number and benefit from increased energy efficiency and better
tunability.

Several recent studies suggest methods for enhancing betatron
radiation emission, mostly based on the increase of the betatron
oscillation amplitude. This can be achieved by an axial magnetic
field, either self-generated or external
\cite{pan2016enhanced,zhang2016enhanced}; by a delayed modulation
laser pulse \cite{lee2019enhanced}; by the interaction of the electron
beam with a high intensity optical lattice formed by the superposition
of two transverse laser pulses \cite{andriyash2013betatron}; by using
structured laser pulses \cite{Martins2019}; or by the interaction
of 
electrons with the tail of the plasma wave drive
pulse\cite{nemeth2008laser,Cipiccia2011,curcio2015resonant,huang2016resonantly}.

The betatron oscillation can also be tuned by manipulation of the
plasma density. This can be done in several ways, e.g.~by using a
tilted shock front in the acceleration phase \cite{yu2018enhanced}, an
axially modulated plasma
density
\cite{palastro2015enhanced}, off-axis laser alignment to a capillary
plasma waveguide \cite{lee2015enhanced}, transverse density gradient
\cite{ferri2018enhancement, kozlova2020hard}, or tailoring the dynamics of the nonlinear plasma wave in a way that electrons find themselves behind its first period (the bubble) for a certain period
of time, where their oscillations are amplified due to the opposite
polarity of transverse fields \cite{mavslarova2019betatron}. Also,
injection of matter by irradiating solid micro-droplets
\cite{yu2014bright} or nanoparticles \cite{chen2013bright} may provide
enhancement of the generated betatron X-ray intensity.

The conversion efficiency from laser-light to X-ray can be increased
by using a hybrid scheme, which combines a low-density laser-driven
plasma accelerator with a high-density beam-driven plasma
radiator\cite{ferri2018high}.  Increase of betatron light by localized
injection of a group of electrons in the shape of an annulus was also
reported \cite{zhao2016high}. The X-ray flux can also be increased due
to shortening of the betatron oscillation wavelength during the natural
longitudinal expansion of bubble \cite{horny2018optical}.

\begin{figure}[htbp]
  \centering
\includegraphics[scale=1.0]{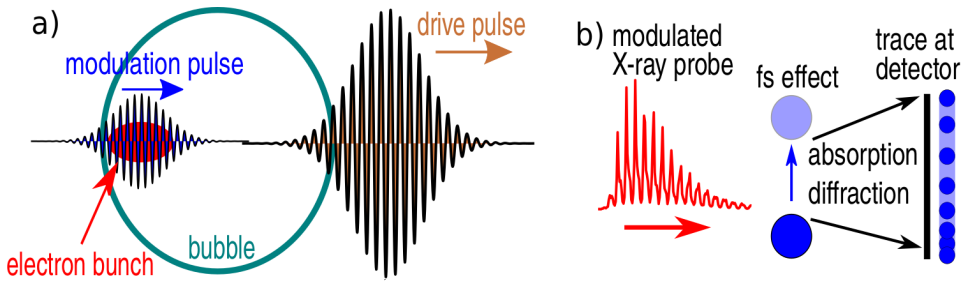} 
\caption{{\bf Schematics of the proposed setup and the application
    configuration}. (a) A moderately high-intensity laser pulse
  creates a plasma cavity free of electrons (bubble). An electron
  bunch is injected in the rear part of the bubble, along with a
  weaker modulation pulse, with a delay that is such that it
  propagates with the electron bunch. (b) Illustration of
  stroboscropic measurement of fast processes using a modulated X-ray
  probe. }
\label{ref:scheme}
\end{figure}

In this paper, we propose  an experimental setup where, in addition to an enhancement of the betatron radiation flux, a train of sub-femtosecond X-ray pulses is generated. It is achieved by separation of the electron bunch accelerated in the laser wakefield into a train of equidistant sub-bunches by a delayed modulation laser pulse, see Figure~\ref{ref:scheme}a) for a schematic of the proposed setup. The separation interval between the pulses corresponds to  half of the modulation pulse wavelength and each pulse in the train is even shorter. 


Generation of electron bunch trains has been studied previously. They originate either from conventional radiofrequency accelerators \cite{petrillo2014dual, shevelev2017generation, dodin2007stochastic}, from laser wakefield accelerators employing  self-injection controlled by driver pulse shaping \cite{kalmykov2018multi} or optical injection by crossing two wakefields\cite{golovin2020generation}, or from plasma wakefield accelerator injected due to the bubble length oscillation on the density downramp \cite{lecz2018trains}.  The advantage of the scheme described in this paper over the aforementioned ones is that the electron bunching is well controlled by the modulator on the sub-micron scale. Thus, the emitted signal comprises of the train of X-ray pulses with an unprecedented repetition rate.

Pulse-trains composed of sub-femtosecond X-ray pulses can enhance the temporal resolution of sampling of ultrafast fundamental physical processes by an order of magnitude, whilst maintaining its other advantageous features such as a small source size of several microns enabling high-resolution images and a relatively small cost of the required laser systems compared to the large scale facilities such as synchrotrons or free electron lasers. A broadband X-ray pulse-train could sample physical processes occurring on femtosecond time-scales by e.g.~X-ray absorption spectroscopy (XAS) or polychromatic (Laue) X-ray diffraction. In all cases, the image observed at the detector (typically a CCD camera) would be composed of a series of sharp and fuzzy regions. As the time-delay between the X-ray pulses in a train is set by the wavelength of the modulation pulse, the dynamics of the sampled process can be extracted from the configuration of the sharp region on the detected image. This approach is analogous to stroboscopic measurement of fast processes, see Figure \ref{ref:scheme}b) for a schematic illustration. In attosecond science, stroboscopic images have been already recorded\cite{mauritsson2008coherent} with high harmonics emission\cite{baltuvska2003attosecond}. Our source, despite being incoherent on its wavelength, provides higher photon energy which results in the increased penetrability through the investigated sample.


\section*{Results}

A driving laser-pulse of moderate intensity
($I \lesssim 10^{19}$~W$\cdot$cm\textsuperscript{-2}), linearly
polarized in the $y$-direction, propagates in the longitudingal ($x$)
direction in an underdense plasma
(in practice, $n_e$ is in the order of $10^{18}$~cm\textsuperscript{-3}) and creates a
moderately nonlinear plasma wave. Its first period, the so-called ``bubble'', is
an ion cavity free of electrons which are expelled by the strong
ponderomotive force of the driving pulse. The electron bunch is
located in the rear part of the bubble. It is injected transversely
($y$-direction), either by self-injection, or as is the case in
this paper, by controlled injection on the density downramp.  A weaker
modulation pulse ($I \lesssim 10^{18}$~W$\cdot$cm\textsuperscript{-2})
with wavelength $\lambda_m$ is injected to follow the driving
pulse. Its electric field, polarized in the $y$-direction, still dominates
over the electrostatic transverse field of the bubble. The delay
between the pulses is chosen in a way that its high-intensity part
co-propagates with the electron bunch.

As the modulation pulse propagates within the bubble, its group
velocity is approximately equal to the speed of light in vacuum
$v_{g,m} \lessapprox c$. The average longitudinal velocity of an
electron in the bunch is lower, due to the relativistic limitation
caused by transverse betatron oscillations. The accelerated electrons
oscillate transversely on a sine-like trajectory because they gained
a considerable transverse momentum dominantly by the fields of the
modulation pulse, but also by the injection process and by the
electrostatic transverse fields of the bubble. Every periodic
increase of their transverse velocity leads to a decrease of their
longitudinal velocity. As a result, the modulation pulse steadily
overtakes the electron bunch.  Consequently, an electron from the
bunch experiences the action of a periodically varying transverse
component of the Lorentz force as it propagates backward with respect
to the modulation pulse.

The transverse electron motion can be described by the equation of motion
$\dif p_y/\dif t \approx q_e(1-\beta_x)E_{0,y,m}\cos(k_m\xi)$,
where $q_e$ is electron charge, $E_{0,y,m}$ is the electric field
amplitude of the modulation pulse, $k_m\xi$ is the phase of the modulation
pulse, with $k_m=2\pi/\lambda_m$ being the modulation pulse wavenumber
and $\xi = x - x_0 - v_{g,m}t$ the coordinate co-moving with the
modulation pulse. Here, we assumed $|p_x| \gg |p_y|$,
$p_x \gg m_ec$, and considered the modulation pulse as a plane wave,
which is applicable in regions around the propagation axis, where its
magnetic field is proportional to its electric field
$B_z \approx E_y/c$.  Thus, the electrons flow backward with respect
to the modulation pulse and due to the phase dependence of the
transverse force, they are periodically pushed in the
$\pm y-$direction. This effect itself leads to enhancement of the
betatron radiation emission in comparison with a standard case without
the modulation pulse.

From the positions where $\cos(k_m\xi) =0$, the absolute value transverse momentum of the electrons decreases and the longitudinal momentum grows; the latter one is largest at the turning points of their trajectory where $p_y=0$. Thus, the turning points related to the modulation pulse phase are the same for all electrons of the bunch. Large longitudinal momenta together with low transverse momenta result in a clustering of the bunch electrons in the nests co-moving with the modulation pulse. Alternatively stated: the original electron bunch is microbunched.  As the betatron radiation is mainly emitted at the turning points of the electron trajectories, its temporal profile is composed of intensity peaks separated by $\lambda_m/2c$, i.e.~a train of X-ray pulses is emitted and the delay between the pulses is adjustable by choosing $\lambda_m$.

The effect of microbunching can be understood as a forced betatron resonance. Contrary to previous cases with the modulation by the tail of the plasma wave drive pulse\cite{Cipiccia2011, huang2016resonantly}, where the electron beam experiences a long acceleration period before it catches the laser pulse which resulting in limited controllability of the X-ray source, we reach the betatron resonance immediately from the moment of injection.

\subsection*{Numerical simulation}
The process of michrobunching and its fingerprint on the betatron radiation signal is studied by means of 2D particle-in-cell (PIC) simulations and their post--processing. A bubble regime configuration with modest laser parameters is chosen for the
demonstration of the process. The parameters used in the simulation
are the following: plasma electron density
$n_0 = 2.5\times 10^{18}$~cm\textsuperscript{-3}, driver laser
wavelength $\lambda_d = 0.8$~$\upmu$m, waist size (radius at
1/e\textsuperscript{2} of maximum intensity) $w_0 = 10$~$\upmu$m,
pulse length (FWHM of intensity) $\tau=20$~fs, and normalized driver laser intensity $a_{0,d} = eE_{0,d}/m_ec \omega_0=1.8$ which corresponds to intensity
$I=6.9\times 10^{18}\;\rm W\;cm^{-2}$. Its focal spot
is located at $x_{f,m}=110$~$\upmu$m. 
The modulation pulse has the same fundamental parameters with the
exception of normalized intensity, which is $a_{0,m}=0.2$, and wavelength
$\lambda_m = \lambda_d/3$ corresponding to intensity
$7.7\times 10^{17}\;\rm W\;cm^{-2}$. It is delayed by
58~fs and its focal spot is located at $x_{f,m}=410$~$\upmu$m. Both
pulses are linearly polarized in the $y-$direction.

Self-injection of electrons in the plasma wakefield does not occur
with these parameters if the plasma density is constant. Instead, a
plasma density profile is chosen so that controlled injection
occurs. In the simulations, the density profile is set in the
following way. A 10~$\upmu$m long vacuum is located at the left edge
of the simulation box, then a 50~$\upmu$m linear density up-ramp
follows until the electron density reaches $2n_e$. Nevertheless,
the nature of the presented injection scheme does not depend on the
plasma-edge density ramp. Afterwards, a 35~$\upmu$m long density
plateau follows; then the density linearly drops to $n_e$ over a
distance of 25~$\upmu$m.  On this down-ramp, the controlled injection
occurs\cite{Geddes}. The PIC simulations were performed with the
\textsc{epoch} code, see the Methods section for details.

\begin{figure*}
  \centering
\includegraphics[scale=0.8]{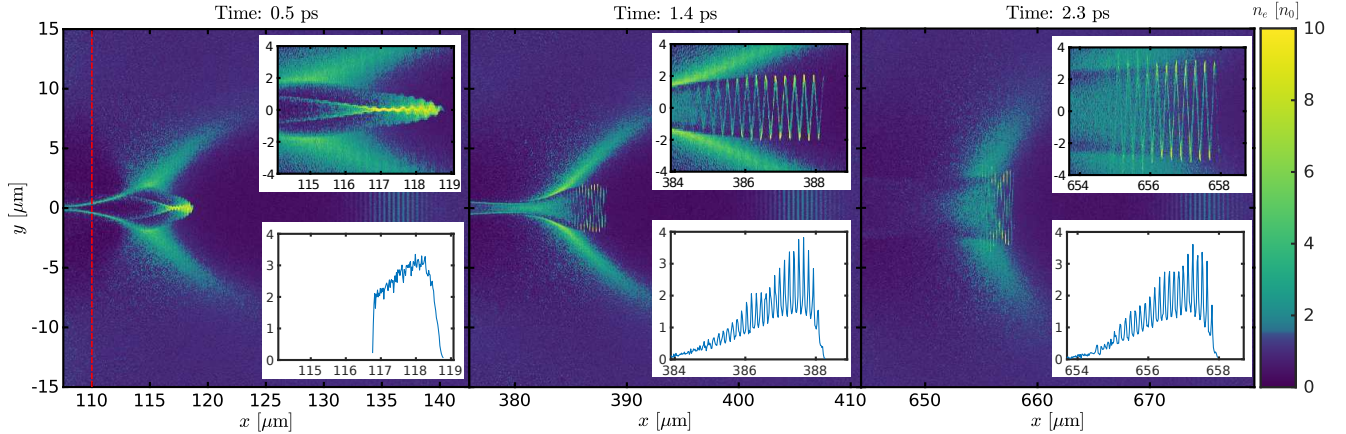}
\caption{{\bf Plasma bubble evolution and electron microbunching.}
  Snapshots of the electron density at the injection time
  ($0.5 \;\rm ps$ left panel) and during the acceleration process
  ($1.4 \;\rm ps$ and $2.3\;\rm ps$, centre and right panels,
  respectively). The red line in the left panel represents the end of
  the initial density down-ramp. Only a central part of the simulation
  box is shown. The upper insets show a zoom of the bunch structure, the bottom insets show a projection of the trapped particles density on the $x-$axis.  }
\label{fig:bubble}
\end{figure*}
The snapshots of the electron density during the injection and
acceleration process are shown in Figure \ref{fig:bubble}. The density
profile in the panel corresponding to the injection time ($t=0.5$~ps)
suggests that the electron bunch is microbunched immediately after the
injection. In later times (1.4~ps and 2.3~ps of simulation), the
snake-like structure of the bunch is pronounced.

\begin{figure}[htbp]
  \centering
\includegraphics[scale=1]{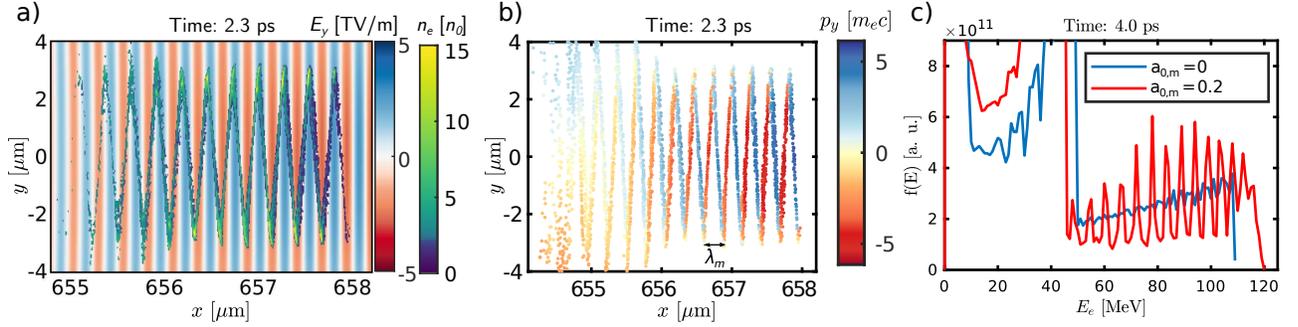}
  \caption{{\bf Electron bunch structure and energy spectrum} a) Electron density of the trapped electrons (plotted
    the simulation cells where average kinetic energy of electrons is
    higher than 10~MeV) and the transverse electric field at
    $t=2.3\;\rm ps$. b) Transverse momentum of the trapped
    electrons. c) Electron energy spectra at
    $t=4.0\;\rm ps$ for the cases with and without modulation pulse
    present. }
\label{fig:DensField}
\end{figure}

The detailed view of the electron bunch structure at 2.3~ps is shown
in Figure \ref{fig:DensField}a), together with the transverse electric field. Apparently, the electric field of the modulation pulse
dominates over the electrostatic field of the bubble in the region
around the axis where the electron bunch is located. The bunch itself
has a sawtooth-shape. The distance between the $x-$coordinates of
the turning points is $\lambda_{m}/2$. The peak values of the electron
density are located in these turning points.

Figure \ref{fig:DensField}b) shows the positions and transverse
momenta of the accelerated electrons. The positions between the peaks
of the density bunch profile and the dominant direction of the
transverse component of the electron momentum confirm that the
electrons propagate backwards in the frame co-moving with the
modulation pulse. These findings can be interpreted as the electron
bunch as a whole performs snake-like motion in the direction of
$-\xi$. This means that the modulation pulse effectively induces the
microbunching of injected electrons and the distance between single
microbunches is $\lambda_{m}/2$ in the longitudinal direction.

The electrons perform betatron oscillations, however, in contrast to
standard betatron motion in the case without the modulation pulse, the
oscillations are driven dominantly by the modulation pulse. Thus, crucially, the
turning points are the same for all of the trapped electrons.
In other words, the electron bunch is effectively separated into
several equidistant microbunches that are continuously
radiating. As a consequence, the observer will receive a modulated betatron radiation signal, comprising of peaks arriving every
$\lambda_m/2c$, as will be shown later.

The electron energy spectrum in time of 4.0~ps just before the structure begins to dephase is shown in Figure \ref{fig:DensField}c); blue and red lines
show the cases without and with the modulation pulse,
respectively. The spectra comprise a clear peak which corresponds to
the electrons accelerated in the first period of the plasma wave due
to the controlled injection. Although, the relative energy spread is
rather high. However, for the purpose of betatron radiation generation
the energy spread is not a determining factor. The presence of the
modulator leads to further electron energy gain compared to the
reference case: the electrons receive the energy stored in the
modulator by direct laser acceleration
\cite{zhangprl2015,shawprl2017}. The estimated accelerated charge
(electron energy higher than 25~MeV) is about 4 to 8~pC in both
cases. There are about 1.3\% less electrons trapped when the modulator
is present.

\subsection*{Betatron radiation spectrogram}
\begin{figure}[htbp]
  \centering
    \includegraphics[scale=1.0]{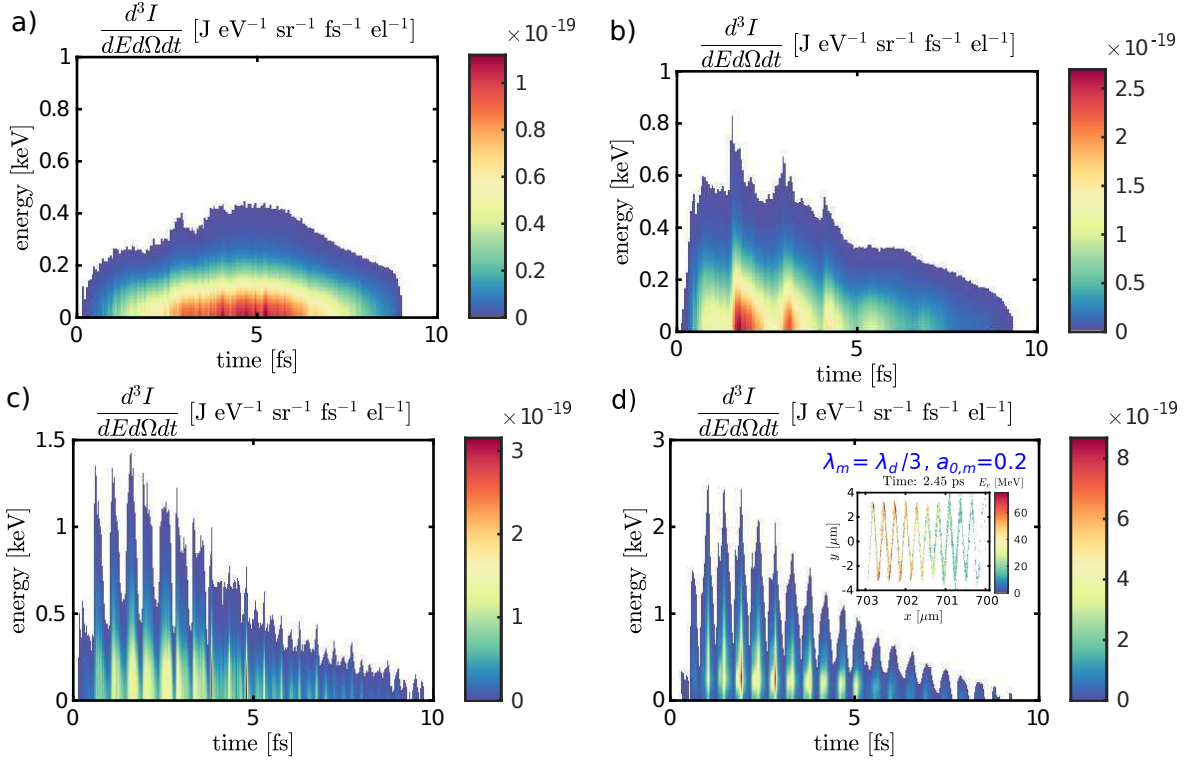}
  \caption{{\bf Spectrograms of the betatron radiation emitted
       by the electrons.} Temporal and energy profiles are shown for a
    reference case without a modulator and for three different
    modulator pulse cases. The signal close to $t = 0$ corresponds to
    the front of the bunch and arrives first at the detector. The
    inset in the panel d) shows the electron energy distribution
    within the bunch. It displays a matrix of the average electron
    energy in cell; only the cells with average energy over 10~MeV are
    shown.  Both temporal and energy profile of emitted X-rays are
    correlated with the inner structure of the bunch. Note that the
    $x$-axis is reversed. }
\label{fig:spec}
\end{figure}

Figure \ref{fig:spec} shows the spectrograms, i.e.~both temporal and
energy profiles of the betatron radiation, with and without the
modulation pulse; for details see the Methods section. Four different
cases are presented: a) the case when the modulator is not present,
(b) with $\lambda_m = \lambda_d$ and $a_{0,m}=0.6$, (c) with
$\lambda_m = \lambda_d/3$ and $a_{0,m}=0.1$, and (d) with $\lambda_m =
\lambda_d/3$ and $a_{0,m}=0.2$. The results presented in
Figs.~\ref{fig:bubble} and \ref{fig:DensField} correspond to  case (d).

All the signals are approximately 10~fs long, corresponding to a bunch
length of $\approx$3.5~$\upmu$m shown in Figure
\ref{fig:DensField}. Nevertheless, while the signal is continuous in
the case without the modulator (Fig.~\ref{fig:spec}a), the modulated
signals (Fig.~\ref{fig:spec}b-d) exhibit trains of ultrashort
pulses. Moreover, the spectrograms show that the betatron radiation critical energy is also modulated in time. In average, the energy of radiation is considerably higher when
the modulator is present. The inset in panel (d) confirms the
correlation between the energy distribution of electrons within the
bunch and the temporal and energy profile of emitted X-rays.

\begin{figure}[htbp]\centering
\includegraphics[scale=1.0]{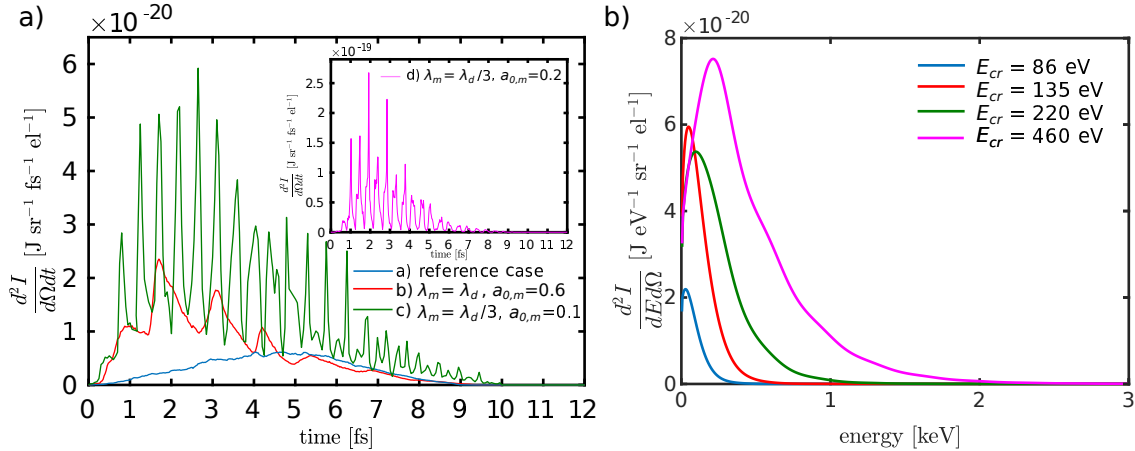} \\
\caption{ {\bf Energy spectra for the emitted betatron radiation}
a) Temporal profile of the betatron radiation for a reference case without a modulator and for three different modulator pulse cases. The inset, corresponding to the case  $\lambda_m = \lambda_d/3$ and $a_{0,m}=0.2$ is an order of magnitude more intense than the other cases. b) On-axis time-integrated energy spectra of emitted X-rays and the critical energy of the emitted signal for the four cases a) no modulator,
(b)  $\lambda_m = \lambda_d$ and $a_{0,m}=0.6$, (c) 
$\lambda_m = \lambda_d/3$ and $a_{0,m}=0.1$, and (d)  $\lambda_m =
\lambda_d/3$ and $a_{0,m}=0.2$. }
\label{fig:inteandint}
\end{figure}

Figure \ref{fig:inteandint} shows the temporal profiles of betatron
radiation. Whereas the blue curve belonging to reference case (a) does
not vary significantly, the other three curves (b-d) show several
clear peaks. The red curve represents the case (b); three dominant
peaks are present. The peak-to-peak distances is between the first and
the second and the second and the third dominant peaks are 1.35~fs and
1.29~fs, respectively. This is in good agreement with the
theoretically expected value $\lambda_m/2c=1.\overline{3}$~fs. The green curve
corresponds to  case (c). The signal comprises of more than thirteen
clear peaks. The peak-to-peak distance is (0.46$\pm$0.02)~fs
(estimated by Fourier transform of signal) and is in good agreement with the expected value of $\lambda_m/2c=0.\overline{4}$~fs. Such a feature can be interpreted as a betatron radiation pulse train coherence with respect to the modulation pulse.

The radiation
peaks themselves are even shorter, the FWHM of the brightest one at
2.65~fs is 140~as. There is a considerable continuous background, the
pulsed signal to noise ratio is about 5:1. This ratio could be
significantly improved by employing a transmission filter which
effectively cuts the low energy parts of the spectra.

The inset of Figure \ref{fig:inteandint}a contains the last case
(d). The signal is an order of a magnitude more intense than the
other cases. It is bunched, with a signal-to-noise ratio of better than
20:1. Again, Fourier transform of this signal shows that the
fundamental period is (0.45$\pm$0.01)~fs, and the FWHM of the brightest peak at 1.92~fs is 100~as.

The number of electrons within the bunch differs by less than 3.5\% between all four compared cases.The estimated total energy within the pulse train is 0.10~nJ in  case (a). It increases greatly when the modulator in present: it is 0.45~nJ, 0.65~nJ, and 2.2 nJ  in cases (b-d), respectively. The increase is caused partly by the higher energy of the electrons and partly by the higher amplitude of betatron oscillations.

Finally, the time-integrated energy spectra on axis for all the cases
(a-d) are shown in Figure \ref{fig:inteandint}b, including information
about the critical energy of the emitted signal in all cases. The critical
energy of the case (d) is 5.3$\times$ higher than in the reference
case (a).

\section*{Discussion}
We propose a method for producing a train of ultrashort X-ray pulses by modifying the standard laser wakefield accelerator setup delivering betatron radiation. This is accomplished by adding a
delayed modulation laser pulse to follow the plasma wave in the region where the electron bunch is injected. As a result, the betatron oscillations of the accelerated electrons are driven dominantly by the fields of the modulation pulse and not by electrostatic fields of the bubble. The turning points of the betatron trajectories are the same for all accelerated electrons and the electrons cluster there.

In other words, the electron bunch is microbunched and the longitudinal distance between the single bunches is half of the modulation pulse wavelength $\lambda_m$.  This property is imprinted on the temporal profile of the emitted X-rays. Thus the betatron radiation signal is composed of a train of pulses separated by a factor of $\lambda_m/2c$, which is 440~as when third harmonics of a standard Ti:sapphire laser pulse is used as the modulator. Moreover, the energy and intensity of the emitted X-rays are also enhanced.  The resulting X-ray source could enable observation of temporal evolution of ultrafast phenomena on the time scale of hundreds of attoseconds.

The process of electron microbunching was further tested in a relatively broad parameter space. The scheme works in the densities $1.8\times 10^{18}$~cm\textsuperscript{-3} -- $6\times 10^{18}$~cm\textsuperscript{-3}. The sharpest microbunching occurs in lower densities, as higher density leads to the lower plasma wave phase velocity causing the structure decay due to dephasing. The results that are presented throughout the paper are given after 3.5~ps of acceleration time ($t=4.0$~ps). This corresponds to the time when the spectrogram is the sharpest for the main demonstration case (d). 

Furthermore, the intensity of the modulator pulse was varied and the stability of the scheme was confirmed. Generally, it can be stated that the scheme works in the parameter range where the modulator pulse field is higher than the transverse electrostatic field of the bubble, but low enough to avoid the significant disruption of the plasma wave. Approximately, this correspond to the normalized modulator pulse intensity of  $a_{0,m} \in [0.05,0.4]$. Within this parameter region, the more intense modulator leads to the better bunching.


We close with two example applications where the suggested technique has the potential to drive forward development.
Betatron radiation has already been used in laboratory astrophysics, when warm dense matter (WDM) samples were investigated employing XAS\cite{mahieu2018probing}. It takes advantage of the broadband photon spectrum in the keV region, where most elements' absorption edges are located. The time-resolved XAS technique pushes its limits from hundreds of picoseconds by synchrotrons or streak cameras to femtoseconds by a betatron source. The presented technique provides an improvement of the XAS time resolution by an order of magnitude.

Broadband synchrotron X-ray pulses are used also in solid state
physics for polychromatic (Laue) X-ray diffraction \cite{li2018situ,
  wu2016intragranular}, where the different energies are diffracted in
different angles. In the standard monochromatic X-ray diffraction,
time resolved synchrotron pulses are used to sample the nonlinear
lattice dynamics, in particular, to determine the crystal structure of
solids and its evolution\cite{mankowsky2014nonlinear,
  buzzi2019measuring}. The pulse train produced by our scheme allows
the development of sub-femtosecond time resolved polychromatic X-ray
diffraction.

\section*{Methods}
2D PIC simulations were performed with the EPOCH \cite{arber2015contemporary} code.  The simulations were run in the moving simulation box with dimensions 80~$\upmu$m $\times$ 40~$\upmu$m. The grid resolution was 90 and 12 cells per $\lambda_d$ in the longitudinal and transverse directions, respectively. Initially, two electron macroparticles were placed in every cell. The plasma is represented as an electron gas; the ions were considered as a homogeneous static background. In total, approximately $2.2\times 10^8$ macroparticles were simulated. 

The temporal profile of betatron radiation was calculated using the method based on the Fourier transform of the emitted signal which can be determined by using trajectories of the trapped electrons\cite{horny2017temporal}. It takes advantage of the fact that each electron performs betatron motion in the wiggler regime and the emitted signal is composed of a series of sharp peaks radiated at the turning points of the electron trajectories separated by relatively long intervals of silence. Thus, it is possible to  store the times when the single peaks of all the tracked electrons were emitted and construct the betatron radiation spectrogram from that. This method is applicable even for the discussed case of X-ray emission by microbunched electrons, because the level of microbunching does not suffice to emit coherent electromagnetic radiation more energetic than ultraviolet.  20~000 of the tracked electron macroparticles were processed in each case.

\section*{Acknowledgments}       
The authors thank 
Václav Petržílka
from IPP CAS, Evangelos Siminos from Gothenburg University, and Julien
Ferri and Longqing Yi from Chalmers University of Technology, for their suggestions and fruitful discussions. This work was supported by the Ministry of Education, Youth and Sports of the Czech Republic within the project LQ1606, from the High Field Initiative (CZ.02.1.01/0.0/0.0/15\_003/0000449) from European Regional Development Fund, and also received funding from the European Research Council (ERC) under the European Union's Horizon 2020 research and innovation programme under grant agreement No 647121.

Access to computing and storage facilities owned by parties and
projects contributing to the National Grid Infrastructure MetaCentrum
provided under the programme \emph{Projects of Large Research,
  Development, and Innovations Infrastructures} (CESNET LM2015042),
and to ECLIPSE cluster of ELI-Beamlines project and at Chalmers Centre
for Computational Science and Engineering (C3SE) provided by the
Swedish National Infrastructure for Computing is greatly appreciated
as well.

\section*{Author contributions statement}
V.H. and W.Y. conceived the idea. V.H. performed the simulations. V.H. and M.K. developed the theoretical interpretation. All authors discussed the findings and contributed to the writing of the manuscript.

\section*{Additional information}
 \textbf{Competing financial interests} The authors declare no competing financial interests. 

\bibliography{report} 

\begin{thebibliography}{54}%
\makeatletter
\providecommand \@ifxundefined [1]{%
 \@ifx{#1\undefined}
}%
\providecommand \@ifnum [1]{%
 \ifnum #1\expandafter \@firstoftwo
 \else \expandafter \@secondoftwo
 \fi
}%
\providecommand \@ifx [1]{%
 \ifx #1\expandafter \@firstoftwo
 \else \expandafter \@secondoftwo
 \fi
}%
\providecommand \natexlab [1]{#1}%
\providecommand \enquote  [1]{``#1''}%
\providecommand \bibnamefont  [1]{#1}%
\providecommand \bibfnamefont [1]{#1}%
\providecommand \citenamefont [1]{#1}%
\providecommand \href@noop [0]{\@secondoftwo}%
\providecommand \href [0]{\begingroup \@sanitize@url \@href}%
\providecommand \@href[1]{\@@startlink{#1}\@@href}%
\providecommand \@@href[1]{\endgroup#1\@@endlink}%
\providecommand \@sanitize@url [0]{\catcode `\\12\catcode `\$12\catcode
  `\&12\catcode `\#12\catcode `\^12\catcode `\_12\catcode `\%12\relax}%
\providecommand \@@startlink[1]{}%
\providecommand \@@endlink[0]{}%
\providecommand \url  [0]{\begingroup\@sanitize@url \@url }%
\providecommand \@url [1]{\endgroup\@href {#1}{\urlprefix }}%
\providecommand \urlprefix  [0]{URL }%
\providecommand \Eprint [0]{\href }%
\providecommand \doibase [0]{http://dx.doi.org/}%
\providecommand \selectlanguage [0]{\@gobble}%
\providecommand \bibinfo  [0]{\@secondoftwo}%
\providecommand \bibfield  [0]{\@secondoftwo}%
\providecommand \translation [1]{[#1]}%
\providecommand \BibitemOpen [0]{}%
\providecommand \bibitemStop [0]{}%
\providecommand \bibitemNoStop [0]{.\EOS\space}%
\providecommand \EOS [0]{\spacefactor3000\relax}%
\providecommand \BibitemShut  [1]{\csname bibitem#1\endcsname}%
\let\auto@bib@innerbib\@empty
\bibitem [{\citenamefont {Martin}\ and\ \citenamefont
  {Hynes}(2004)}]{martin2004femtochemistry}%
  \BibitemOpen
  \bibfield  {author} {\bibinfo {author} {\bibfnamefont {M.~M.}\ \bibnamefont
  {Martin}}\ and\ \bibinfo {author} {\bibfnamefont {J.~T.}\ \bibnamefont
  {Hynes}},\ }\href@noop {} {\emph {\bibinfo {title} {Femtochemistry and
  femtobiology: ultrafast events in molecular science}}}\ (\bibinfo
  {publisher} {Elsevier},\ \bibinfo {year} {2004})\BibitemShut {NoStop}%
\bibitem [{\citenamefont {G{\"o}tzfried}\ \emph {et~al.}(2018)\citenamefont
  {G{\"o}tzfried}, \citenamefont {D{\"o}pp}, \citenamefont {Gilljohann},
  \citenamefont {Ding}, \citenamefont {Schindler}, \citenamefont {Wenz},
  \citenamefont {Hehn}, \citenamefont {Pfeiffer},\ and\ \citenamefont
  {Karsch}}]{gotzfried2018research}%
  \BibitemOpen
  \bibfield  {author} {\bibinfo {author} {\bibfnamefont {J.}~\bibnamefont
  {G{\"o}tzfried}}, \bibinfo {author} {\bibfnamefont {A.}~\bibnamefont
  {D{\"o}pp}}, \bibinfo {author} {\bibfnamefont {M.}~\bibnamefont
  {Gilljohann}}, \bibinfo {author} {\bibfnamefont {H.}~\bibnamefont {Ding}},
  \bibinfo {author} {\bibfnamefont {S.}~\bibnamefont {Schindler}}, \bibinfo
  {author} {\bibfnamefont {J.}~\bibnamefont {Wenz}}, \bibinfo {author}
  {\bibfnamefont {L.}~\bibnamefont {Hehn}}, \bibinfo {author} {\bibfnamefont
  {F.}~\bibnamefont {Pfeiffer}}, \ and\ \bibinfo {author} {\bibfnamefont
  {S.}~\bibnamefont {Karsch}},\ }\href {\doibase
  https://doi.org/10.1016/j.nima.2018.02.110} {\bibfield  {journal} {\bibinfo
  {journal} {Nuclear Instruments and Methods in Physics Research Section A:
  Accelerators, Spectrometers, Detectors and Associated Equipment}\ }\textbf
  {\bibinfo {volume} {909}},\ \bibinfo {pages} {286} (\bibinfo {year}
  {2018})}\BibitemShut {NoStop}%
\bibitem [{\citenamefont {Bilderback}\ \emph {et~al.}(2005)\citenamefont
  {Bilderback}, \citenamefont {Elleaume},\ and\ \citenamefont
  {Weckert}}]{bilderback2005review}%
  \BibitemOpen
  \bibfield  {author} {\bibinfo {author} {\bibfnamefont {D.~H.}\ \bibnamefont
  {Bilderback}}, \bibinfo {author} {\bibfnamefont {P.}~\bibnamefont
  {Elleaume}}, \ and\ \bibinfo {author} {\bibfnamefont {E.}~\bibnamefont
  {Weckert}},\ }\href {\doibase http://dx.doi.org/10.1088/0953-4075/38/9/022}
  {\bibfield  {journal} {\bibinfo  {journal} {{Journal of Physics B: Atomic,
  Molecular and Optical Physics}}\ }\textbf {\bibinfo {volume} {38}},\ \bibinfo
  {pages} {S773} (\bibinfo {year} {2005})}\BibitemShut {NoStop}%
\bibitem [{\citenamefont {McNeil}\ and\ \citenamefont
  {Thompson}(2010)}]{mcneil2010x}%
  \BibitemOpen
  \bibfield  {author} {\bibinfo {author} {\bibfnamefont {B.~W.}\ \bibnamefont
  {McNeil}}\ and\ \bibinfo {author} {\bibfnamefont {N.~R.}\ \bibnamefont
  {Thompson}},\ }\href {\doibase http://dx.doi.org/10.1038/nphoton.2010.239}
  {\bibfield  {journal} {\bibinfo  {journal} {Nature Photonics}\ }\textbf
  {\bibinfo {volume} {4}},\ \bibinfo {pages} {814} (\bibinfo {year}
  {2010})}\BibitemShut {NoStop}%
\bibitem [{\citenamefont {Esarey}\ \emph {et~al.}(2009)\citenamefont {Esarey},
  \citenamefont {Schroeder},\ and\ \citenamefont
  {Leemans}}]{esarey2009physics}%
  \BibitemOpen
  \bibfield  {author} {\bibinfo {author} {\bibfnamefont {E.}~\bibnamefont
  {Esarey}}, \bibinfo {author} {\bibfnamefont {C.~B.}\ \bibnamefont
  {Schroeder}}, \ and\ \bibinfo {author} {\bibfnamefont {W.~P.}\ \bibnamefont
  {Leemans}},\ }\href {\doibase https://doi.org/10.1103/RevModPhys.81.1229}
  {\bibfield  {journal} {\bibinfo  {journal} {Reviews of Modern Physics}\
  }\textbf {\bibinfo {volume} {81}},\ \bibinfo {pages} {1229} (\bibinfo {year}
  {2009})}\BibitemShut {NoStop}%
\bibitem [{\citenamefont {Gonsalves}\ \emph {et~al.}(2019)\citenamefont
  {Gonsalves}, \citenamefont {Nakamura}, \citenamefont {Daniels}, \citenamefont
  {Benedetti}, \citenamefont {Pieronek}, \citenamefont {de~Raadt},
  \citenamefont {Steinke}, \citenamefont {Bin}, \citenamefont {Bulanov},
  \citenamefont {van Tilborg}, \citenamefont {Geddes}, \citenamefont
  {Schroeder}, \citenamefont {T\'oth}, \citenamefont {Esarey}, \citenamefont
  {Swanson}, \citenamefont {Fan-Chiang}, \citenamefont {Bagdasarov},
  \citenamefont {Bobrova}, \citenamefont {Gasilov}, \citenamefont {Korn},
  \citenamefont {Sasorov},\ and\ \citenamefont
  {Leemans}}]{gonsalves2019petawatt}%
  \BibitemOpen
  \bibfield  {author} {\bibinfo {author} {\bibfnamefont {A.~J.}\ \bibnamefont
  {Gonsalves}}, \bibinfo {author} {\bibfnamefont {K.}~\bibnamefont {Nakamura}},
  \bibinfo {author} {\bibfnamefont {J.}~\bibnamefont {Daniels}}, \bibinfo
  {author} {\bibfnamefont {C.}~\bibnamefont {Benedetti}}, \bibinfo {author}
  {\bibfnamefont {C.}~\bibnamefont {Pieronek}}, \bibinfo {author}
  {\bibfnamefont {T.~C.~H.}\ \bibnamefont {de~Raadt}}, \bibinfo {author}
  {\bibfnamefont {S.}~\bibnamefont {Steinke}}, \bibinfo {author} {\bibfnamefont
  {J.~H.}\ \bibnamefont {Bin}}, \bibinfo {author} {\bibfnamefont {S.~S.}\
  \bibnamefont {Bulanov}}, \bibinfo {author} {\bibfnamefont {J.}~\bibnamefont
  {van Tilborg}}, \bibinfo {author} {\bibfnamefont {C.~G.~R.}\ \bibnamefont
  {Geddes}}, \bibinfo {author} {\bibfnamefont {C.~B.}\ \bibnamefont
  {Schroeder}}, \bibinfo {author} {\bibfnamefont {C.}~\bibnamefont {T\'oth}},
  \bibinfo {author} {\bibfnamefont {E.}~\bibnamefont {Esarey}}, \bibinfo
  {author} {\bibfnamefont {K.}~\bibnamefont {Swanson}}, \bibinfo {author}
  {\bibfnamefont {L.}~\bibnamefont {Fan-Chiang}}, \bibinfo {author}
  {\bibfnamefont {G.}~\bibnamefont {Bagdasarov}}, \bibinfo {author}
  {\bibfnamefont {N.}~\bibnamefont {Bobrova}}, \bibinfo {author} {\bibfnamefont
  {V.}~\bibnamefont {Gasilov}}, \bibinfo {author} {\bibfnamefont
  {G.}~\bibnamefont {Korn}}, \bibinfo {author} {\bibfnamefont {P.}~\bibnamefont
  {Sasorov}}, \ and\ \bibinfo {author} {\bibfnamefont {W.~P.}\ \bibnamefont
  {Leemans}},\ }\href {\doibase
  http://dx.doi.org/10.1103/PhysRevLett.122.084801} {\bibfield  {journal}
  {\bibinfo  {journal} {Physical Review Letters}\ }\textbf {\bibinfo {volume}
  {122}},\ \bibinfo {pages} {084801} (\bibinfo {year} {2019})}\BibitemShut
  {NoStop}%
\bibitem [{\citenamefont {Blumenfeld}\ \emph {et~al.}(2007)\citenamefont
  {Blumenfeld}, \citenamefont {Clayton}, \citenamefont {Decker}, \citenamefont
  {Hogan}, \citenamefont {Huang}, \citenamefont {Ischebeck}, \citenamefont
  {Iverson}, \citenamefont {Joshi}, \citenamefont {Katsouleas}, \citenamefont
  {Kirby} \emph {et~al.}}]{blumenfeld2007energy}%
  \BibitemOpen
  \bibfield  {author} {\bibinfo {author} {\bibfnamefont {I.}~\bibnamefont
  {Blumenfeld}}, \bibinfo {author} {\bibfnamefont {C.~E.}\ \bibnamefont
  {Clayton}}, \bibinfo {author} {\bibfnamefont {F.-J.}\ \bibnamefont {Decker}},
  \bibinfo {author} {\bibfnamefont {M.~J.}\ \bibnamefont {Hogan}}, \bibinfo
  {author} {\bibfnamefont {C.}~\bibnamefont {Huang}}, \bibinfo {author}
  {\bibfnamefont {R.}~\bibnamefont {Ischebeck}}, \bibinfo {author}
  {\bibfnamefont {R.}~\bibnamefont {Iverson}}, \bibinfo {author} {\bibfnamefont
  {C.}~\bibnamefont {Joshi}}, \bibinfo {author} {\bibfnamefont
  {T.}~\bibnamefont {Katsouleas}}, \bibinfo {author} {\bibfnamefont
  {N.}~\bibnamefont {Kirby}},  \emph {et~al.},\ }\href {\doibase
  http://dx.doi.org/10.1038/nature05538} {\bibfield  {journal} {\bibinfo
  {journal} {Nature}\ }\textbf {\bibinfo {volume} {445}},\ \bibinfo {pages}
  {741} (\bibinfo {year} {2007})}\BibitemShut {NoStop}%
\bibitem [{\citenamefont {Adli}\ \emph {et~al.}(2018)\citenamefont {Adli},
  \citenamefont {Ahuja}, \citenamefont {Apsimon}, \citenamefont {Apsimon},
  \citenamefont {Bachmann}, \citenamefont {Barrientos}, \citenamefont {Batsch},
  \citenamefont {Bauche}, \citenamefont {Olsen}, \citenamefont {Bernardini}
  \emph {et~al.}}]{adli2018acceleration}%
  \BibitemOpen
  \bibfield  {author} {\bibinfo {author} {\bibfnamefont {E.}~\bibnamefont
  {Adli}}, \bibinfo {author} {\bibfnamefont {A.}~\bibnamefont {Ahuja}},
  \bibinfo {author} {\bibfnamefont {O.}~\bibnamefont {Apsimon}}, \bibinfo
  {author} {\bibfnamefont {R.}~\bibnamefont {Apsimon}}, \bibinfo {author}
  {\bibfnamefont {A.-M.}\ \bibnamefont {Bachmann}}, \bibinfo {author}
  {\bibfnamefont {D.}~\bibnamefont {Barrientos}}, \bibinfo {author}
  {\bibfnamefont {F.}~\bibnamefont {Batsch}}, \bibinfo {author} {\bibfnamefont
  {J.}~\bibnamefont {Bauche}}, \bibinfo {author} {\bibfnamefont {V.~B.}\
  \bibnamefont {Olsen}}, \bibinfo {author} {\bibfnamefont {M.}~\bibnamefont
  {Bernardini}},  \emph {et~al.},\ }\href {\doibase
  http://dx.doi.org/10.1038/s41586-018-0485-4} {\bibfield  {journal} {\bibinfo
  {journal} {Nature}\ }\textbf {\bibinfo {volume} {561}},\ \bibinfo {pages}
  {363} (\bibinfo {year} {2018})}\BibitemShut {NoStop}%
\bibitem [{\citenamefont {Kiselev}\ \emph {et~al.}(2004)\citenamefont
  {Kiselev}, \citenamefont {Pukhov},\ and\ \citenamefont
  {Kostyukov}}]{kiselev2004x}%
  \BibitemOpen
  \bibfield  {author} {\bibinfo {author} {\bibfnamefont {S.}~\bibnamefont
  {Kiselev}}, \bibinfo {author} {\bibfnamefont {A.}~\bibnamefont {Pukhov}}, \
  and\ \bibinfo {author} {\bibfnamefont {I.}~\bibnamefont {Kostyukov}},\ }\href
  {\doibase https://doi.org/10.1103/PhysRevLett.93.135004} {\bibfield
  {journal} {\bibinfo  {journal} {Physical Review Letters}\ }\textbf {\bibinfo
  {volume} {93}},\ \bibinfo {pages} {135004} (\bibinfo {year}
  {2004})}\BibitemShut {NoStop}%
\bibitem [{\citenamefont {Rousse}\ \emph {et~al.}(2004)\citenamefont {Rousse},
  \citenamefont {Phuoc}, \citenamefont {Shah}, \citenamefont {Pukhov},
  \citenamefont {Lefebvre}, \citenamefont {Malka}, \citenamefont {Kiselev},
  \citenamefont {Burgy}, \citenamefont {Rousseau}, \citenamefont {Umstadter}
  \emph {et~al.}}]{rousse2004production}%
  \BibitemOpen
  \bibfield  {author} {\bibinfo {author} {\bibfnamefont {A.}~\bibnamefont
  {Rousse}}, \bibinfo {author} {\bibfnamefont {K.~T.}\ \bibnamefont {Phuoc}},
  \bibinfo {author} {\bibfnamefont {R.}~\bibnamefont {Shah}}, \bibinfo {author}
  {\bibfnamefont {A.}~\bibnamefont {Pukhov}}, \bibinfo {author} {\bibfnamefont
  {E.}~\bibnamefont {Lefebvre}}, \bibinfo {author} {\bibfnamefont
  {V.}~\bibnamefont {Malka}}, \bibinfo {author} {\bibfnamefont
  {S.}~\bibnamefont {Kiselev}}, \bibinfo {author} {\bibfnamefont
  {F.}~\bibnamefont {Burgy}}, \bibinfo {author} {\bibfnamefont {J.-P.}\
  \bibnamefont {Rousseau}}, \bibinfo {author} {\bibfnamefont {D.}~\bibnamefont
  {Umstadter}},  \emph {et~al.},\ }\href {\doibase
  https://doi.org/10.1103/PhysRevLett.93.135005} {\bibfield  {journal}
  {\bibinfo  {journal} {Physical Review Letters}\ }\textbf {\bibinfo {volume}
  {93}},\ \bibinfo {pages} {135005} (\bibinfo {year} {2004})}\BibitemShut
  {NoStop}%
\bibitem [{\citenamefont {Schnell}\ \emph {et~al.}(2015)\citenamefont
  {Schnell}, \citenamefont {S{\"a}vert}, \citenamefont {Uschmann},
  \citenamefont {Jansen}, \citenamefont {Kaluza},\ and\ \citenamefont
  {Spielmann}}]{schnell2015characterization}%
  \BibitemOpen
  \bibfield  {author} {\bibinfo {author} {\bibfnamefont {M.}~\bibnamefont
  {Schnell}}, \bibinfo {author} {\bibfnamefont {A.}~\bibnamefont {S{\"a}vert}},
  \bibinfo {author} {\bibfnamefont {I.}~\bibnamefont {Uschmann}}, \bibinfo
  {author} {\bibfnamefont {O.}~\bibnamefont {Jansen}}, \bibinfo {author}
  {\bibfnamefont {M.~C.}\ \bibnamefont {Kaluza}}, \ and\ \bibinfo {author}
  {\bibfnamefont {C.}~\bibnamefont {Spielmann}},\ }\href {\doibase
  https://doi.org/10.1017/S0022377815000379} {\bibfield  {journal} {\bibinfo
  {journal} {Journal of Plasma Physics}\ }\textbf {\bibinfo {volume} {81}}
  (\bibinfo {year} {2015}),\
  https://doi.org/10.1017/S0022377815000379}\BibitemShut {NoStop}%
\bibitem [{\citenamefont {Fourmaux}\ \emph {et~al.}(2011)\citenamefont
  {Fourmaux}, \citenamefont {Corde}, \citenamefont {Phuoc}, \citenamefont
  {Leguay}, \citenamefont {Payeur}, \citenamefont {Lassonde}, \citenamefont
  {Gnedyuk}, \citenamefont {Lebrun}, \citenamefont {Fourment}, \citenamefont
  {Malka} \emph {et~al.}}]{fourmaux2011demonstration}%
  \BibitemOpen
  \bibfield  {author} {\bibinfo {author} {\bibfnamefont {S.}~\bibnamefont
  {Fourmaux}}, \bibinfo {author} {\bibfnamefont {S.}~\bibnamefont {Corde}},
  \bibinfo {author} {\bibfnamefont {K.~T.}\ \bibnamefont {Phuoc}}, \bibinfo
  {author} {\bibfnamefont {P.}~\bibnamefont {Leguay}}, \bibinfo {author}
  {\bibfnamefont {S.}~\bibnamefont {Payeur}}, \bibinfo {author} {\bibfnamefont
  {P.}~\bibnamefont {Lassonde}}, \bibinfo {author} {\bibfnamefont
  {S.}~\bibnamefont {Gnedyuk}}, \bibinfo {author} {\bibfnamefont
  {G.}~\bibnamefont {Lebrun}}, \bibinfo {author} {\bibfnamefont
  {C.}~\bibnamefont {Fourment}}, \bibinfo {author} {\bibfnamefont
  {V.}~\bibnamefont {Malka}},  \emph {et~al.},\ }\href {\doibase
  https://doi.org/10.1088/1367-2630/13/3/033017} {\bibfield  {journal}
  {\bibinfo  {journal} {New Journal of Physics}\ }\textbf {\bibinfo {volume}
  {13}},\ \bibinfo {pages} {033017} (\bibinfo {year} {2011})}\BibitemShut
  {NoStop}%
\bibitem [{\citenamefont {Rousse}\ \emph {et~al.}(2007)\citenamefont {Rousse},
  \citenamefont {Phuoc}, \citenamefont {Shah}, \citenamefont {Fitour},\ and\
  \citenamefont {Albert}}]{rousse2007scaling}%
  \BibitemOpen
  \bibfield  {author} {\bibinfo {author} {\bibfnamefont {A.}~\bibnamefont
  {Rousse}}, \bibinfo {author} {\bibfnamefont {K.~T.}\ \bibnamefont {Phuoc}},
  \bibinfo {author} {\bibfnamefont {R.}~\bibnamefont {Shah}}, \bibinfo {author}
  {\bibfnamefont {R.}~\bibnamefont {Fitour}}, \ and\ \bibinfo {author}
  {\bibfnamefont {F.}~\bibnamefont {Albert}},\ }\href {\doibase
  10.1140/epjd/e2007-00249-7} {\bibfield  {journal} {\bibinfo  {journal} {The
  European Physical Journal D}\ }\textbf {\bibinfo {volume} {45}},\ \bibinfo
  {pages} {391} (\bibinfo {year} {2007})}\BibitemShut {NoStop}%
\bibitem [{\citenamefont {Corde}\ \emph {et~al.}(2013)\citenamefont {Corde},
  \citenamefont {Phuoc}, \citenamefont {Lambert}, \citenamefont {Fitour},
  \citenamefont {Malka}, \citenamefont {Rousse}, \citenamefont {Beck},\ and\
  \citenamefont {Lefebvre}}]{corde2013femtosecond}%
  \BibitemOpen
  \bibfield  {author} {\bibinfo {author} {\bibfnamefont {S.}~\bibnamefont
  {Corde}}, \bibinfo {author} {\bibfnamefont {K.~T.}\ \bibnamefont {Phuoc}},
  \bibinfo {author} {\bibfnamefont {G.}~\bibnamefont {Lambert}}, \bibinfo
  {author} {\bibfnamefont {R.}~\bibnamefont {Fitour}}, \bibinfo {author}
  {\bibfnamefont {V.}~\bibnamefont {Malka}}, \bibinfo {author} {\bibfnamefont
  {A.}~\bibnamefont {Rousse}}, \bibinfo {author} {\bibfnamefont
  {A.}~\bibnamefont {Beck}}, \ and\ \bibinfo {author} {\bibfnamefont
  {E.}~\bibnamefont {Lefebvre}},\ }\href {\doibase
  https://doi.org/10.1103/RevModPhys.85.1} {\bibfield  {journal} {\bibinfo
  {journal} {Reviews of Modern Physics}\ }\textbf {\bibinfo {volume} {85}},\
  \bibinfo {pages} {1} (\bibinfo {year} {2013})}\BibitemShut {NoStop}%
\bibitem [{\citenamefont {Cole}\ \emph {et~al.}(2015)\citenamefont {Cole},
  \citenamefont {Wood}, \citenamefont {Lopes}, \citenamefont {Poder},
  \citenamefont {Abel}, \citenamefont {Alatabi}, \citenamefont {Bryant},
  \citenamefont {Jin}, \citenamefont {Kneip}, \citenamefont {Mecseki},
  \citenamefont {Symes}, \citenamefont {Mangles},\ and\ \citenamefont
  {Najmudin}}]{Cole2015}%
  \BibitemOpen
  \bibfield  {author} {\bibinfo {author} {\bibfnamefont {J.~M.}\ \bibnamefont
  {Cole}}, \bibinfo {author} {\bibfnamefont {J.~C.}\ \bibnamefont {Wood}},
  \bibinfo {author} {\bibfnamefont {N.~C.}\ \bibnamefont {Lopes}}, \bibinfo
  {author} {\bibfnamefont {K.}~\bibnamefont {Poder}}, \bibinfo {author}
  {\bibfnamefont {R.~L.}\ \bibnamefont {Abel}}, \bibinfo {author}
  {\bibfnamefont {S.}~\bibnamefont {Alatabi}}, \bibinfo {author} {\bibfnamefont
  {J.~S.~J.}\ \bibnamefont {Bryant}}, \bibinfo {author} {\bibfnamefont
  {A.}~\bibnamefont {Jin}}, \bibinfo {author} {\bibfnamefont {S.}~\bibnamefont
  {Kneip}}, \bibinfo {author} {\bibfnamefont {K.}~\bibnamefont {Mecseki}},
  \bibinfo {author} {\bibfnamefont {D.~R.}\ \bibnamefont {Symes}}, \bibinfo
  {author} {\bibfnamefont {S.~P.~D.}\ \bibnamefont {Mangles}}, \ and\ \bibinfo
  {author} {\bibfnamefont {Z.}~\bibnamefont {Najmudin}},\ }\href {\doibase
  https://doi.org/10.1038/srep13244} {\bibfield  {journal} {\bibinfo  {journal}
  {Scientific Reports}\ }\textbf {\bibinfo {volume} {5}},\ \bibinfo {pages}
  {13244 EP } (\bibinfo {year} {2015})}\BibitemShut {NoStop}%
\bibitem [{\citenamefont {Albert}\ and\ \citenamefont
  {Thomas}(2016)}]{Albert_2016}%
  \BibitemOpen
  \bibfield  {author} {\bibinfo {author} {\bibfnamefont {F.}~\bibnamefont
  {Albert}}\ and\ \bibinfo {author} {\bibfnamefont {A.~G.~R.}\ \bibnamefont
  {Thomas}},\ }\href {\doibase 10.1088/0741-3335/58/10/103001} {\bibfield
  {journal} {\bibinfo  {journal} {Plasma Physics and Controlled Fusion}\
  }\textbf {\bibinfo {volume} {58}},\ \bibinfo {pages} {103001} (\bibinfo
  {year} {2016})}\BibitemShut {NoStop}%
\bibitem [{\citenamefont {Pan}\ \emph {et~al.}(2016)\citenamefont {Pan},
  \citenamefont {Zheng}, \citenamefont {Cao}, \citenamefont {Liu},\ and\
  \citenamefont {He}}]{pan2016enhanced}%
  \BibitemOpen
  \bibfield  {author} {\bibinfo {author} {\bibfnamefont {K.}~\bibnamefont
  {Pan}}, \bibinfo {author} {\bibfnamefont {C.}~\bibnamefont {Zheng}}, \bibinfo
  {author} {\bibfnamefont {L.}~\bibnamefont {Cao}}, \bibinfo {author}
  {\bibfnamefont {Z.}~\bibnamefont {Liu}}, \ and\ \bibinfo {author}
  {\bibfnamefont {X.}~\bibnamefont {He}},\ }\href {\doibase
  https://doi.org/10.1063/1.4947545} {\bibfield  {journal} {\bibinfo  {journal}
  {Physics of Plasmas}\ }\textbf {\bibinfo {volume} {23}},\ \bibinfo {pages}
  {043115} (\bibinfo {year} {2016})}\BibitemShut {NoStop}%
\bibitem [{\citenamefont {Zhang}\ \emph {et~al.}(2016)\citenamefont {Zhang},
  \citenamefont {Zhang}, \citenamefont {Hong}, \citenamefont {Yu},
  \citenamefont {Deng}, \citenamefont {Teng}, \citenamefont {He},\ and\
  \citenamefont {Gu}}]{zhang2016enhanced}%
  \BibitemOpen
  \bibfield  {author} {\bibinfo {author} {\bibfnamefont {Z.}~\bibnamefont
  {Zhang}}, \bibinfo {author} {\bibfnamefont {B.}~\bibnamefont {Zhang}},
  \bibinfo {author} {\bibfnamefont {W.}~\bibnamefont {Hong}}, \bibinfo {author}
  {\bibfnamefont {M.}~\bibnamefont {Yu}}, \bibinfo {author} {\bibfnamefont
  {Z.}~\bibnamefont {Deng}}, \bibinfo {author} {\bibfnamefont {J.}~\bibnamefont
  {Teng}}, \bibinfo {author} {\bibfnamefont {S.}~\bibnamefont {He}}, \ and\
  \bibinfo {author} {\bibfnamefont {Y.}~\bibnamefont {Gu}},\ }\href {\doibase
  http://dx.doi.org/10.1088/0741-3335/58/10/105009} {\bibfield  {journal}
  {\bibinfo  {journal} {Plasma Physics and Controlled Fusion}\ }\textbf
  {\bibinfo {volume} {58}},\ \bibinfo {pages} {105009} (\bibinfo {year}
  {2016})}\BibitemShut {NoStop}%
\bibitem [{\citenamefont {Lee}\ \emph {et~al.}(2019)\citenamefont {Lee},
  \citenamefont {Uhm}, \citenamefont {Kang}, \citenamefont {Hur},\ and\
  \citenamefont {Suk}}]{lee2019enhanced}%
  \BibitemOpen
  \bibfield  {author} {\bibinfo {author} {\bibfnamefont {S.}~\bibnamefont
  {Lee}}, \bibinfo {author} {\bibfnamefont {H.~S.}\ \bibnamefont {Uhm}},
  \bibinfo {author} {\bibfnamefont {T.~Y.}\ \bibnamefont {Kang}}, \bibinfo
  {author} {\bibfnamefont {M.~S.}\ \bibnamefont {Hur}}, \ and\ \bibinfo
  {author} {\bibfnamefont {H.}~\bibnamefont {Suk}},\ }\href {\doibase
  https://doi.org/10.1016/j.cap.2019.01.018} {\bibfield  {journal} {\bibinfo
  {journal} {Current Applied Physics}\ }\textbf {\bibinfo {volume} {19}},\
  \bibinfo {pages} {464} (\bibinfo {year} {2019})}\BibitemShut {NoStop}%
\bibitem [{\citenamefont {Andriyash}\ \emph {et~al.}(2013)\citenamefont
  {Andriyash}, \citenamefont {d’Humi{\`e}res}, \citenamefont {Tikhonchuk},\
  and\ \citenamefont {Balcou}}]{andriyash2013betatron}%
  \BibitemOpen
  \bibfield  {author} {\bibinfo {author} {\bibfnamefont {I.}~\bibnamefont
  {Andriyash}}, \bibinfo {author} {\bibfnamefont {E.}~\bibnamefont
  {d’Humi{\`e}res}}, \bibinfo {author} {\bibfnamefont {V.}~\bibnamefont
  {Tikhonchuk}}, \ and\ \bibinfo {author} {\bibfnamefont {P.}~\bibnamefont
  {Balcou}},\ }\href {\doibase https://doi.org/10.1103/PhysRevSTAB.16.100703}
  {\bibfield  {journal} {\bibinfo  {journal} {Physical Review Special
  Topics-Accelerators and Beams}\ }\textbf {\bibinfo {volume} {16}},\ \bibinfo
  {pages} {100703} (\bibinfo {year} {2013})}\BibitemShut {NoStop}%
\bibitem [{\citenamefont {Lu{\'i}s~Martins}\ \emph {et~al.}(2019)\citenamefont
  {Lu{\'i}s~Martins}, \citenamefont {Vieira}, \citenamefont {Ferri},\ and\
  \citenamefont {F{\"u}l{\"o}p}}]{Martins2019}%
  \BibitemOpen
  \bibfield  {author} {\bibinfo {author} {\bibfnamefont {J.}~\bibnamefont
  {Lu{\'i}s~Martins}}, \bibinfo {author} {\bibfnamefont {J.}~\bibnamefont
  {Vieira}}, \bibinfo {author} {\bibfnamefont {J.}~\bibnamefont {Ferri}}, \
  and\ \bibinfo {author} {\bibfnamefont {T.}~\bibnamefont {F{\"u}l{\"o}p}},\
  }\href {\doibase 10.1038/s41598-019-45474-8} {\bibfield  {journal} {\bibinfo
  {journal} {Scientific Reports}\ }\textbf {\bibinfo {volume} {9}},\ \bibinfo
  {pages} {9840} (\bibinfo {year} {2019})}\BibitemShut {NoStop}%
\bibitem [{\citenamefont {Németh}\ \emph {et~al.}(2008)\citenamefont
  {Németh}, \citenamefont {Shen}, \citenamefont {Li}, \citenamefont {Shang},
  \citenamefont {Crowell}, \citenamefont {Harkay},\ and\ \citenamefont
  {Cary}}]{nemeth2008laser}%
  \BibitemOpen
  \bibfield  {author} {\bibinfo {author} {\bibfnamefont {K.}~\bibnamefont
  {Németh}}, \bibinfo {author} {\bibfnamefont {B.}~\bibnamefont {Shen}},
  \bibinfo {author} {\bibfnamefont {Y.}~\bibnamefont {Li}}, \bibinfo {author}
  {\bibfnamefont {H.}~\bibnamefont {Shang}}, \bibinfo {author} {\bibfnamefont
  {R.}~\bibnamefont {Crowell}}, \bibinfo {author} {\bibfnamefont {K.~C.}\
  \bibnamefont {Harkay}}, \ and\ \bibinfo {author} {\bibfnamefont {J.~R.}\
  \bibnamefont {Cary}},\ }\href {\doibase
  https://doi.org/10.1103/PhysRevLett.100.095002} {\bibfield  {journal}
  {\bibinfo  {journal} {Physical Review Letters}\ }\textbf {\bibinfo {volume}
  {100}},\ \bibinfo {pages} {095002} (\bibinfo {year} {2008})}\BibitemShut
  {NoStop}%
\bibitem [{\citenamefont {Cipiccia}\ \emph {et~al.}(2011)\citenamefont
  {Cipiccia}, \citenamefont {Islam}, \citenamefont {Ersfeld}, \citenamefont
  {Shanks}, \citenamefont {Brunetti}, \citenamefont {Vieux}, \citenamefont
  {Yang}, \citenamefont {Issac}, \citenamefont {Wiggins}, \citenamefont
  {Welsh}, \citenamefont {Anania}, \citenamefont {Maneuski}, \citenamefont
  {Montgomery}, \citenamefont {Smith}, \citenamefont {Hoek}, \citenamefont
  {Hamilton}, \citenamefont {Lemos}, \citenamefont {Symes}, \citenamefont
  {Rajeev}, \citenamefont {Shea}, \citenamefont {Dias},\ and\ \citenamefont
  {Jaroszynski}}]{Cipiccia2011}%
  \BibitemOpen
  \bibfield  {author} {\bibinfo {author} {\bibfnamefont {S.}~\bibnamefont
  {Cipiccia}}, \bibinfo {author} {\bibfnamefont {M.~R.}\ \bibnamefont {Islam}},
  \bibinfo {author} {\bibfnamefont {B.}~\bibnamefont {Ersfeld}}, \bibinfo
  {author} {\bibfnamefont {R.~P.}\ \bibnamefont {Shanks}}, \bibinfo {author}
  {\bibfnamefont {E.}~\bibnamefont {Brunetti}}, \bibinfo {author}
  {\bibfnamefont {G.}~\bibnamefont {Vieux}}, \bibinfo {author} {\bibfnamefont
  {X.}~\bibnamefont {Yang}}, \bibinfo {author} {\bibfnamefont {R.~C.}\
  \bibnamefont {Issac}}, \bibinfo {author} {\bibfnamefont {S.~M.}\ \bibnamefont
  {Wiggins}}, \bibinfo {author} {\bibfnamefont {G.~H.}\ \bibnamefont {Welsh}},
  \bibinfo {author} {\bibfnamefont {M.-P.}\ \bibnamefont {Anania}}, \bibinfo
  {author} {\bibfnamefont {D.}~\bibnamefont {Maneuski}}, \bibinfo {author}
  {\bibfnamefont {R.}~\bibnamefont {Montgomery}}, \bibinfo {author}
  {\bibfnamefont {G.}~\bibnamefont {Smith}}, \bibinfo {author} {\bibfnamefont
  {M.}~\bibnamefont {Hoek}}, \bibinfo {author} {\bibfnamefont {D.~J.}\
  \bibnamefont {Hamilton}}, \bibinfo {author} {\bibfnamefont {N.~R.~C.}\
  \bibnamefont {Lemos}}, \bibinfo {author} {\bibfnamefont {D.}~\bibnamefont
  {Symes}}, \bibinfo {author} {\bibfnamefont {P.~P.}\ \bibnamefont {Rajeev}},
  \bibinfo {author} {\bibfnamefont {V.~O.}\ \bibnamefont {Shea}}, \bibinfo
  {author} {\bibfnamefont {J.~M.}\ \bibnamefont {Dias}}, \ and\ \bibinfo
  {author} {\bibfnamefont {D.~A.}\ \bibnamefont {Jaroszynski}},\ }\href
  {\doibase https://doi.org/10.1038/nphys2090} {\bibfield  {journal} {\bibinfo
  {journal} {Nature Physics}\ }\textbf {\bibinfo {volume} {7}},\ \bibinfo
  {pages} {867} (\bibinfo {year} {2011})}\BibitemShut {NoStop}%
\bibitem [{\citenamefont {Curcio}\ \emph {et~al.}(2015)\citenamefont {Curcio},
  \citenamefont {Giulietti}, \citenamefont {Dattoli},\ and\ \citenamefont
  {Ferrario}}]{curcio2015resonant}%
  \BibitemOpen
  \bibfield  {author} {\bibinfo {author} {\bibfnamefont {A.}~\bibnamefont
  {Curcio}}, \bibinfo {author} {\bibfnamefont {D.}~\bibnamefont {Giulietti}},
  \bibinfo {author} {\bibfnamefont {G.}~\bibnamefont {Dattoli}}, \ and\
  \bibinfo {author} {\bibfnamefont {M.}~\bibnamefont {Ferrario}},\ }\href
  {\doibase https://doi.org/10.1017/S0022377815000926} {\bibfield  {journal}
  {\bibinfo  {journal} {Journal of Plasma Physics}\ }\textbf {\bibinfo {volume}
  {81}} (\bibinfo {year} {2015}),\
  https://doi.org/10.1017/S0022377815000926}\BibitemShut {NoStop}%
\bibitem [{\citenamefont {Huang}\ \emph {et~al.}(2016)\citenamefont {Huang},
  \citenamefont {Li}, \citenamefont {Li}, \citenamefont {Chen}, \citenamefont
  {Tao}, \citenamefont {Ma}, \citenamefont {Zhao}, \citenamefont {Li},
  \citenamefont {Chen}, \citenamefont {Mirzaie} \emph
  {et~al.}}]{huang2016resonantly}%
  \BibitemOpen
  \bibfield  {author} {\bibinfo {author} {\bibfnamefont {K.}~\bibnamefont
  {Huang}}, \bibinfo {author} {\bibfnamefont {Y.}~\bibnamefont {Li}}, \bibinfo
  {author} {\bibfnamefont {D.}~\bibnamefont {Li}}, \bibinfo {author}
  {\bibfnamefont {L.}~\bibnamefont {Chen}}, \bibinfo {author} {\bibfnamefont
  {M.}~\bibnamefont {Tao}}, \bibinfo {author} {\bibfnamefont {Y.}~\bibnamefont
  {Ma}}, \bibinfo {author} {\bibfnamefont {J.}~\bibnamefont {Zhao}}, \bibinfo
  {author} {\bibfnamefont {M.}~\bibnamefont {Li}}, \bibinfo {author}
  {\bibfnamefont {M.}~\bibnamefont {Chen}}, \bibinfo {author} {\bibfnamefont
  {M.}~\bibnamefont {Mirzaie}},  \emph {et~al.},\ }\href {\doibase
  https://doi.org/10.1038/srep27633} {\bibfield  {journal} {\bibinfo  {journal}
  {Scientific Reports}\ }\textbf {\bibinfo {volume} {6}},\ \bibinfo {pages}
  {27633} (\bibinfo {year} {2016})}\BibitemShut {NoStop}%
\bibitem [{\citenamefont {Yu}\ \emph {et~al.}(2018)\citenamefont {Yu},
  \citenamefont {Liu}, \citenamefont {Wang}, \citenamefont {Li}, \citenamefont
  {Qi}, \citenamefont {Zhang}, \citenamefont {Qin}, \citenamefont {Liu},
  \citenamefont {Fang}, \citenamefont {Feng} \emph {et~al.}}]{yu2018enhanced}%
  \BibitemOpen
  \bibfield  {author} {\bibinfo {author} {\bibfnamefont {C.}~\bibnamefont
  {Yu}}, \bibinfo {author} {\bibfnamefont {J.}~\bibnamefont {Liu}}, \bibinfo
  {author} {\bibfnamefont {W.}~\bibnamefont {Wang}}, \bibinfo {author}
  {\bibfnamefont {W.}~\bibnamefont {Li}}, \bibinfo {author} {\bibfnamefont
  {R.}~\bibnamefont {Qi}}, \bibinfo {author} {\bibfnamefont {Z.}~\bibnamefont
  {Zhang}}, \bibinfo {author} {\bibfnamefont {Z.}~\bibnamefont {Qin}}, \bibinfo
  {author} {\bibfnamefont {J.}~\bibnamefont {Liu}}, \bibinfo {author}
  {\bibfnamefont {M.}~\bibnamefont {Fang}}, \bibinfo {author} {\bibfnamefont
  {K.}~\bibnamefont {Feng}},  \emph {et~al.},\ }\href {\doibase
  https://doi.org/10.1063/1.5019406} {\bibfield  {journal} {\bibinfo  {journal}
  {Applied Physics Letters}\ }\textbf {\bibinfo {volume} {112}},\ \bibinfo
  {pages} {133503} (\bibinfo {year} {2018})}\BibitemShut {NoStop}%
\bibitem [{\citenamefont {Palastro}\ \emph {et~al.}(2015)\citenamefont
  {Palastro}, \citenamefont {Kaganovich},\ and\ \citenamefont
  {Gordon}}]{palastro2015enhanced}%
  \BibitemOpen
  \bibfield  {author} {\bibinfo {author} {\bibfnamefont {J.}~\bibnamefont
  {Palastro}}, \bibinfo {author} {\bibfnamefont {D.}~\bibnamefont
  {Kaganovich}}, \ and\ \bibinfo {author} {\bibfnamefont {D.}~\bibnamefont
  {Gordon}},\ }\href {\doibase https://doi.org/10.1063/1.4923018} {\bibfield
  {journal} {\bibinfo  {journal} {Physics of Plasmas}\ }\textbf {\bibinfo
  {volume} {22}},\ \bibinfo {pages} {063111} (\bibinfo {year}
  {2015})}\BibitemShut {NoStop}%
\bibitem [{\citenamefont {Lee}\ \emph {et~al.}(2015)\citenamefont {Lee},
  \citenamefont {Lee}, \citenamefont {Gupta}, \citenamefont {Uhm},\ and\
  \citenamefont {Suk}}]{lee2015enhanced}%
  \BibitemOpen
  \bibfield  {author} {\bibinfo {author} {\bibfnamefont {S.}~\bibnamefont
  {Lee}}, \bibinfo {author} {\bibfnamefont {T.}~\bibnamefont {Lee}}, \bibinfo
  {author} {\bibfnamefont {D.}~\bibnamefont {Gupta}}, \bibinfo {author}
  {\bibfnamefont {H.}~\bibnamefont {Uhm}}, \ and\ \bibinfo {author}
  {\bibfnamefont {H.}~\bibnamefont {Suk}},\ }\href {\doibase
  http://dx.doi.org/10.1088/0741-3335/57/7/075002} {\bibfield  {journal}
  {\bibinfo  {journal} {Plasma Physics and Controlled Fusion}\ }\textbf
  {\bibinfo {volume} {57}},\ \bibinfo {pages} {075002} (\bibinfo {year}
  {2015})}\BibitemShut {NoStop}%
\bibitem [{\citenamefont {Ferri}\ and\ \citenamefont
  {Davoine}(2018)}]{ferri2018enhancement}%
  \BibitemOpen
  \bibfield  {author} {\bibinfo {author} {\bibfnamefont {J.}~\bibnamefont
  {Ferri}}\ and\ \bibinfo {author} {\bibfnamefont {X.}~\bibnamefont
  {Davoine}},\ }\href {\doibase
  https://doi.org/10.1103/PhysRevAccelBeams.21.091302} {\bibfield  {journal}
  {\bibinfo  {journal} {Physical Review Accelerators and Beams}\ }\textbf
  {\bibinfo {volume} {21}},\ \bibinfo {pages} {091302} (\bibinfo {year}
  {2018})}\BibitemShut {NoStop}%
\bibitem [{\citenamefont {Kozlová}\ \emph {et~al.}(2020)\citenamefont
  {Kozlová}, \citenamefont {Andriyash}, \citenamefont {Gautier}, \citenamefont
  {Sebban}, \citenamefont {Smartsev}, \citenamefont {Jourdain}, \citenamefont
  {Chulagain}, \citenamefont {Azamoum}, \citenamefont {Tafzi}, \citenamefont
  {Goddet} \emph {et~al.}}]{kozlova2020hard}%
  \BibitemOpen
  \bibfield  {author} {\bibinfo {author} {\bibfnamefont {M.}~\bibnamefont
  {Kozlová}}, \bibinfo {author} {\bibfnamefont {I.}~\bibnamefont {Andriyash}},
  \bibinfo {author} {\bibfnamefont {J.}~\bibnamefont {Gautier}}, \bibinfo
  {author} {\bibfnamefont {S.}~\bibnamefont {Sebban}}, \bibinfo {author}
  {\bibfnamefont {S.}~\bibnamefont {Smartsev}}, \bibinfo {author}
  {\bibfnamefont {N.}~\bibnamefont {Jourdain}}, \bibinfo {author}
  {\bibfnamefont {U.}~\bibnamefont {Chulagain}}, \bibinfo {author}
  {\bibfnamefont {Y.}~\bibnamefont {Azamoum}}, \bibinfo {author} {\bibfnamefont
  {A.}~\bibnamefont {Tafzi}}, \bibinfo {author} {\bibfnamefont {J.-P.}\
  \bibnamefont {Goddet}},  \emph {et~al.},\ }\href {\doibase
  https://doi.org/10.1103/PhysRevX.10.011061} {\bibfield  {journal} {\bibinfo
  {journal} {Physical Review X}\ }\textbf {\bibinfo {volume} {10}},\ \bibinfo
  {pages} {011061} (\bibinfo {year} {2020})}\BibitemShut {NoStop}%
\bibitem [{\citenamefont {Mašlárová}\ \emph {et~al.}(2019)\citenamefont
  {Mašlárová}, \citenamefont {Horný}, \citenamefont {Krůs},\ and\
  \citenamefont {Pšikal}}]{mavslarova2019betatron}%
  \BibitemOpen
  \bibfield  {author} {\bibinfo {author} {\bibfnamefont {D.}~\bibnamefont
  {Mašlárová}}, \bibinfo {author} {\bibfnamefont {V.}~\bibnamefont
  {Horný}}, \bibinfo {author} {\bibfnamefont {M.}~\bibnamefont {Krůs}}, \
  and\ \bibinfo {author} {\bibfnamefont {J.}~\bibnamefont {Pšikal}},\ }in\
  \href {\doibase https://doi.org/10.1117/12.2520980} {\emph {\bibinfo
  {booktitle} {Laser Acceleration of Electrons, Protons, and Ions V}}},\ Vol.\
  \bibinfo {volume} {11037}\ (\bibinfo {organization} {International Society
  for Optics and Photonics},\ \bibinfo {year} {2019})\ p.\ \bibinfo {pages}
  {1103710}\BibitemShut {NoStop}%
\bibitem [{\citenamefont {Yu}\ \emph {et~al.}(2014)\citenamefont {Yu},
  \citenamefont {Hu}, \citenamefont {Yin}, \citenamefont {Shao}, \citenamefont
  {Zhuo}, \citenamefont {Ma}, \citenamefont {Yang}, \citenamefont {Luo},\ and\
  \citenamefont {Pukhov}}]{yu2014bright}%
  \BibitemOpen
  \bibfield  {author} {\bibinfo {author} {\bibfnamefont {T.-P.}\ \bibnamefont
  {Yu}}, \bibinfo {author} {\bibfnamefont {L.-X.}\ \bibnamefont {Hu}}, \bibinfo
  {author} {\bibfnamefont {Y.}~\bibnamefont {Yin}}, \bibinfo {author}
  {\bibfnamefont {F.-Q.}\ \bibnamefont {Shao}}, \bibinfo {author}
  {\bibfnamefont {H.-B.}\ \bibnamefont {Zhuo}}, \bibinfo {author}
  {\bibfnamefont {Y.-Y.}\ \bibnamefont {Ma}}, \bibinfo {author} {\bibfnamefont
  {X.-H.}\ \bibnamefont {Yang}}, \bibinfo {author} {\bibfnamefont
  {W.}~\bibnamefont {Luo}}, \ and\ \bibinfo {author} {\bibfnamefont
  {A.}~\bibnamefont {Pukhov}},\ }\href {\doibase
  https://doi.org/10.1063/1.4895928} {\bibfield  {journal} {\bibinfo  {journal}
  {Applied Physics Letters}\ }\textbf {\bibinfo {volume} {105}},\ \bibinfo
  {pages} {114101} (\bibinfo {year} {2014})}\BibitemShut {NoStop}%
\bibitem [{\citenamefont {Chen}\ \emph {et~al.}(2013)\citenamefont {Chen},
  \citenamefont {Yan}, \citenamefont {Li}, \citenamefont {Hu}, \citenamefont
  {Zhang}, \citenamefont {Wang}, \citenamefont {Hafz}, \citenamefont {Mao},
  \citenamefont {Huang}, \citenamefont {Ma} \emph {et~al.}}]{chen2013bright}%
  \BibitemOpen
  \bibfield  {author} {\bibinfo {author} {\bibfnamefont {L.}~\bibnamefont
  {Chen}}, \bibinfo {author} {\bibfnamefont {W.}~\bibnamefont {Yan}}, \bibinfo
  {author} {\bibfnamefont {D.}~\bibnamefont {Li}}, \bibinfo {author}
  {\bibfnamefont {Z.}~\bibnamefont {Hu}}, \bibinfo {author} {\bibfnamefont
  {L.}~\bibnamefont {Zhang}}, \bibinfo {author} {\bibfnamefont
  {W.}~\bibnamefont {Wang}}, \bibinfo {author} {\bibfnamefont {N.}~\bibnamefont
  {Hafz}}, \bibinfo {author} {\bibfnamefont {J.}~\bibnamefont {Mao}}, \bibinfo
  {author} {\bibfnamefont {K.}~\bibnamefont {Huang}}, \bibinfo {author}
  {\bibfnamefont {Y.}~\bibnamefont {Ma}},  \emph {et~al.},\ }\href {\doibase
  https://doi.org/10.1038/srep01912} {\bibfield  {journal} {\bibinfo  {journal}
  {Scientific Reports}\ }\textbf {\bibinfo {volume} {3}},\ \bibinfo {pages}
  {1912} (\bibinfo {year} {2013})}\BibitemShut {NoStop}%
\bibitem [{\citenamefont {Ferri}\ \emph {et~al.}(2018)\citenamefont {Ferri},
  \citenamefont {Corde}, \citenamefont {D{\"o}pp}, \citenamefont {Lifschitz},
  \citenamefont {Doche}, \citenamefont {Thaury}, \citenamefont {Phuoc},
  \citenamefont {Mahieu}, \citenamefont {Andriyash}, \citenamefont {Malka}
  \emph {et~al.}}]{ferri2018high}%
  \BibitemOpen
  \bibfield  {author} {\bibinfo {author} {\bibfnamefont {J.}~\bibnamefont
  {Ferri}}, \bibinfo {author} {\bibfnamefont {S.}~\bibnamefont {Corde}},
  \bibinfo {author} {\bibfnamefont {A.}~\bibnamefont {D{\"o}pp}}, \bibinfo
  {author} {\bibfnamefont {A.}~\bibnamefont {Lifschitz}}, \bibinfo {author}
  {\bibfnamefont {A.}~\bibnamefont {Doche}}, \bibinfo {author} {\bibfnamefont
  {C.}~\bibnamefont {Thaury}}, \bibinfo {author} {\bibfnamefont {K.~T.}\
  \bibnamefont {Phuoc}}, \bibinfo {author} {\bibfnamefont {B.}~\bibnamefont
  {Mahieu}}, \bibinfo {author} {\bibfnamefont {I.}~\bibnamefont {Andriyash}},
  \bibinfo {author} {\bibfnamefont {V.}~\bibnamefont {Malka}},  \emph
  {et~al.},\ }\href {\doibase https://doi.org/10.1103/PhysRevLett.120.254802}
  {\bibfield  {journal} {\bibinfo  {journal} {Physical Review Letters}\
  }\textbf {\bibinfo {volume} {120}},\ \bibinfo {pages} {254802} (\bibinfo
  {year} {2018})}\BibitemShut {NoStop}%
\bibitem [{\citenamefont {Zhao}\ \emph {et~al.}(2016)\citenamefont {Zhao},
  \citenamefont {Behm}, \citenamefont {Dong}, \citenamefont {Davoine},
  \citenamefont {Kalmykov}, \citenamefont {Petrov}, \citenamefont {Chvykov},
  \citenamefont {Cummings}, \citenamefont {Hou}, \citenamefont {Maksimchuk}
  \emph {et~al.}}]{zhao2016high}%
  \BibitemOpen
  \bibfield  {author} {\bibinfo {author} {\bibfnamefont {T.}~\bibnamefont
  {Zhao}}, \bibinfo {author} {\bibfnamefont {K.}~\bibnamefont {Behm}}, \bibinfo
  {author} {\bibfnamefont {C.}~\bibnamefont {Dong}}, \bibinfo {author}
  {\bibfnamefont {X.}~\bibnamefont {Davoine}}, \bibinfo {author} {\bibfnamefont
  {S.~Y.}\ \bibnamefont {Kalmykov}}, \bibinfo {author} {\bibfnamefont
  {V.}~\bibnamefont {Petrov}}, \bibinfo {author} {\bibfnamefont
  {V.}~\bibnamefont {Chvykov}}, \bibinfo {author} {\bibfnamefont
  {P.}~\bibnamefont {Cummings}}, \bibinfo {author} {\bibfnamefont
  {B.}~\bibnamefont {Hou}}, \bibinfo {author} {\bibfnamefont {A.}~\bibnamefont
  {Maksimchuk}},  \emph {et~al.},\ }\href
  {https://doi.org/10.1103/PhysRevLett.117.094801} {\bibfield  {journal}
  {\bibinfo  {journal} {Physical Review Letters}\ }\textbf {\bibinfo {volume}
  {117}},\ \bibinfo {pages} {094801} (\bibinfo {year} {2016})}\BibitemShut
  {NoStop}%
\bibitem [{\citenamefont {Horný}\ \emph {et~al.}(2018)\citenamefont {Horný},
  \citenamefont {Mašlárová}, \citenamefont {Petržílka}, \citenamefont
  {Klimo}, \citenamefont {Kozlová},\ and\ \citenamefont
  {Krůs}}]{horny2018optical}%
  \BibitemOpen
  \bibfield  {author} {\bibinfo {author} {\bibfnamefont {V.}~\bibnamefont
  {Horný}}, \bibinfo {author} {\bibfnamefont {D.}~\bibnamefont
  {Mašlárová}}, \bibinfo {author} {\bibfnamefont {V.}~\bibnamefont
  {Petržílka}}, \bibinfo {author} {\bibfnamefont {O.}~\bibnamefont {Klimo}},
  \bibinfo {author} {\bibfnamefont {M.}~\bibnamefont {Kozlová}}, \ and\
  \bibinfo {author} {\bibfnamefont {M.}~\bibnamefont {Krůs}},\ }\href
  {\doibase https://doi.org/10.1088/1361-6587/aabd07} {\bibfield  {journal}
  {\bibinfo  {journal} {Plasma Physics and Controlled Fusion}\ }\textbf
  {\bibinfo {volume} {60}},\ \bibinfo {pages} {064009} (\bibinfo {year}
  {2018})}\BibitemShut {NoStop}%
\bibitem [{\citenamefont {Petrillo}\ \emph {et~al.}(2014)\citenamefont
  {Petrillo}, \citenamefont {Bacci}, \citenamefont {Curatolo}, \citenamefont
  {Ferrario}, \citenamefont {Gatti}, \citenamefont {Maroli}, \citenamefont
  {Rau}, \citenamefont {Ronsivalle}, \citenamefont {Serafini}, \citenamefont
  {Vaccarezza} \emph {et~al.}}]{petrillo2014dual}%
  \BibitemOpen
  \bibfield  {author} {\bibinfo {author} {\bibfnamefont {V.}~\bibnamefont
  {Petrillo}}, \bibinfo {author} {\bibfnamefont {A.}~\bibnamefont {Bacci}},
  \bibinfo {author} {\bibfnamefont {C.}~\bibnamefont {Curatolo}}, \bibinfo
  {author} {\bibfnamefont {M.}~\bibnamefont {Ferrario}}, \bibinfo {author}
  {\bibfnamefont {G.}~\bibnamefont {Gatti}}, \bibinfo {author} {\bibfnamefont
  {C.}~\bibnamefont {Maroli}}, \bibinfo {author} {\bibfnamefont
  {J.}~\bibnamefont {Rau}}, \bibinfo {author} {\bibfnamefont {C.}~\bibnamefont
  {Ronsivalle}}, \bibinfo {author} {\bibfnamefont {L.}~\bibnamefont
  {Serafini}}, \bibinfo {author} {\bibfnamefont {C.}~\bibnamefont
  {Vaccarezza}},  \emph {et~al.},\ }\href {\doibase
  https://doi.org/10.1103/PhysRevSTAB.17.020706} {\bibfield  {journal}
  {\bibinfo  {journal} {Physical Review Special Topics-Accelerators and Beams}\
  }\textbf {\bibinfo {volume} {17}},\ \bibinfo {pages} {020706} (\bibinfo
  {year} {2014})}\BibitemShut {NoStop}%
\bibitem [{\citenamefont {Shevelev}\ \emph {et~al.}(2017)\citenamefont
  {Shevelev}, \citenamefont {Aryshev}, \citenamefont {Terunuma},\ and\
  \citenamefont {Urakawa}}]{shevelev2017generation}%
  \BibitemOpen
  \bibfield  {author} {\bibinfo {author} {\bibfnamefont {M.}~\bibnamefont
  {Shevelev}}, \bibinfo {author} {\bibfnamefont {A.}~\bibnamefont {Aryshev}},
  \bibinfo {author} {\bibfnamefont {N.}~\bibnamefont {Terunuma}}, \ and\
  \bibinfo {author} {\bibfnamefont {J.}~\bibnamefont {Urakawa}},\ }\href
  {\doibase https://doi.org/10.1103/PhysRevAccelBeams.20.103401} {\bibfield
  {journal} {\bibinfo  {journal} {Physical Review Accelerators and Beams}\
  }\textbf {\bibinfo {volume} {20}},\ \bibinfo {pages} {103401} (\bibinfo
  {year} {2017})}\BibitemShut {NoStop}%
\bibitem [{\citenamefont {Dodin}\ and\ \citenamefont
  {Fisch}(2007)}]{dodin2007stochastic}%
  \BibitemOpen
  \bibfield  {author} {\bibinfo {author} {\bibfnamefont {I.}~\bibnamefont
  {Dodin}}\ and\ \bibinfo {author} {\bibfnamefont {N.~J.}\ \bibnamefont
  {Fisch}},\ }\href {\doibase https://doi.org/10.1103/PhysRevLett.98.234801}
  {\bibfield  {journal} {\bibinfo  {journal} {Physical Review Letters}\
  }\textbf {\bibinfo {volume} {98}},\ \bibinfo {pages} {234801} (\bibinfo
  {year} {2007})}\BibitemShut {NoStop}%
\bibitem [{\citenamefont {Kalmykov}\ \emph {et~al.}(2018)\citenamefont
  {Kalmykov}, \citenamefont {Davoine}, \citenamefont {Ghebregziabher},\ and\
  \citenamefont {Shadwick}}]{kalmykov2018multi}%
  \BibitemOpen
  \bibfield  {author} {\bibinfo {author} {\bibfnamefont {S.~Y.}\ \bibnamefont
  {Kalmykov}}, \bibinfo {author} {\bibfnamefont {X.}~\bibnamefont {Davoine}},
  \bibinfo {author} {\bibfnamefont {I.}~\bibnamefont {Ghebregziabher}}, \ and\
  \bibinfo {author} {\bibfnamefont {B.~A.}\ \bibnamefont {Shadwick}},\ }\href
  {\doibase https://doi.org/10.1016/j.nima.2018.02.001} {\bibfield  {journal}
  {\bibinfo  {journal} {Nuclear Instruments and Methods in Physics Research,
  Section A: Accelerators, Spectrometers, Detectors and Associated Equipment}\
  }\textbf {\bibinfo {volume} {909}},\ \bibinfo {pages} {433} (\bibinfo {year}
  {2018})}\BibitemShut {NoStop}%
\bibitem [{\citenamefont {Golovin}\ \emph {et~al.}(2020)\citenamefont
  {Golovin}, \citenamefont {Horný}, \citenamefont {Yan}, \citenamefont
  {Fruhling}, \citenamefont {Haden}, \citenamefont {Wang}, \citenamefont
  {Banerjee},\ and\ \citenamefont {Umstadter}}]{golovin2020generation}%
  \BibitemOpen
  \bibfield  {author} {\bibinfo {author} {\bibfnamefont {G.}~\bibnamefont
  {Golovin}}, \bibinfo {author} {\bibfnamefont {V.}~\bibnamefont {Horný}},
  \bibinfo {author} {\bibfnamefont {W.}~\bibnamefont {Yan}}, \bibinfo {author}
  {\bibfnamefont {C.}~\bibnamefont {Fruhling}}, \bibinfo {author}
  {\bibfnamefont {D.}~\bibnamefont {Haden}}, \bibinfo {author} {\bibfnamefont
  {J.}~\bibnamefont {Wang}}, \bibinfo {author} {\bibfnamefont {S.}~\bibnamefont
  {Banerjee}}, \ and\ \bibinfo {author} {\bibfnamefont {D.}~\bibnamefont
  {Umstadter}},\ }\href {\doibase https://doi.org/10.1063/1.5141953} {\bibfield
   {journal} {\bibinfo  {journal} {Physics of Plasmas}\ }\textbf {\bibinfo
  {volume} {27}},\ \bibinfo {pages} {033105} (\bibinfo {year}
  {2020})}\BibitemShut {NoStop}%
\bibitem [{\citenamefont {L{\'e}cz}\ \emph {et~al.}(2018)\citenamefont
  {L{\'e}cz}, \citenamefont {Andreev}, \citenamefont {Konoplev}, \citenamefont
  {Seryi},\ and\ \citenamefont {Smith}}]{lecz2018trains}%
  \BibitemOpen
  \bibfield  {author} {\bibinfo {author} {\bibfnamefont {Z.}~\bibnamefont
  {L{\'e}cz}}, \bibinfo {author} {\bibfnamefont {A.}~\bibnamefont {Andreev}},
  \bibinfo {author} {\bibfnamefont {I.}~\bibnamefont {Konoplev}}, \bibinfo
  {author} {\bibfnamefont {A.}~\bibnamefont {Seryi}}, \ and\ \bibinfo {author}
  {\bibfnamefont {J.}~\bibnamefont {Smith}},\ }\href {\doibase
  https://doi.org/10.1088/1361-6587/aac064} {\bibfield  {journal} {\bibinfo
  {journal} {Plasma Physics and Controlled Fusion}\ }\textbf {\bibinfo {volume}
  {60}},\ \bibinfo {pages} {075012} (\bibinfo {year} {2018})}\BibitemShut
  {NoStop}%
\bibitem [{\citenamefont {Mauritsson}\ \emph {et~al.}(2008)\citenamefont
  {Mauritsson}, \citenamefont {Johnsson}, \citenamefont {Mansten},
  \citenamefont {Swoboda}, \citenamefont {Ruchon}, \citenamefont
  {L’Huillier},\ and\ \citenamefont {Schafer}}]{mauritsson2008coherent}%
  \BibitemOpen
  \bibfield  {author} {\bibinfo {author} {\bibfnamefont {J.}~\bibnamefont
  {Mauritsson}}, \bibinfo {author} {\bibfnamefont {P.}~\bibnamefont
  {Johnsson}}, \bibinfo {author} {\bibfnamefont {E.}~\bibnamefont {Mansten}},
  \bibinfo {author} {\bibfnamefont {M.}~\bibnamefont {Swoboda}}, \bibinfo
  {author} {\bibfnamefont {T.}~\bibnamefont {Ruchon}}, \bibinfo {author}
  {\bibfnamefont {A.}~\bibnamefont {L’Huillier}}, \ and\ \bibinfo {author}
  {\bibfnamefont {K.}~\bibnamefont {Schafer}},\ }\href {\doibase
  https://doi.org/10.1103/PhysRevLett.100.073003} {\bibfield  {journal}
  {\bibinfo  {journal} {Physical Review Letters}\ }\textbf {\bibinfo {volume}
  {100}},\ \bibinfo {pages} {073003} (\bibinfo {year} {2008})}\BibitemShut
  {NoStop}%
\bibitem [{\citenamefont {Baltuška}\ \emph {et~al.}(2003)\citenamefont
  {Baltuška}, \citenamefont {Udem}, \citenamefont {Uiberacker}, \citenamefont
  {Hentschel}, \citenamefont {Goulielmakis}, \citenamefont {Gohle},
  \citenamefont {Holzwarth}, \citenamefont {Yakovlev}, \citenamefont {Scrinzi},
  \citenamefont {Hänsch} \emph {et~al.}}]{baltuvska2003attosecond}%
  \BibitemOpen
  \bibfield  {author} {\bibinfo {author} {\bibfnamefont {A.}~\bibnamefont
  {Baltuška}}, \bibinfo {author} {\bibfnamefont {T.}~\bibnamefont {Udem}},
  \bibinfo {author} {\bibfnamefont {M.}~\bibnamefont {Uiberacker}}, \bibinfo
  {author} {\bibfnamefont {M.}~\bibnamefont {Hentschel}}, \bibinfo {author}
  {\bibfnamefont {E.}~\bibnamefont {Goulielmakis}}, \bibinfo {author}
  {\bibfnamefont {C.}~\bibnamefont {Gohle}}, \bibinfo {author} {\bibfnamefont
  {R.}~\bibnamefont {Holzwarth}}, \bibinfo {author} {\bibfnamefont
  {V.}~\bibnamefont {Yakovlev}}, \bibinfo {author} {\bibfnamefont
  {A.}~\bibnamefont {Scrinzi}}, \bibinfo {author} {\bibfnamefont {T.~W.}\
  \bibnamefont {Hänsch}},  \emph {et~al.},\ }\href {\doibase
  https://doi.org/10.1038/nature01414} {\bibfield  {journal} {\bibinfo
  {journal} {Nature}\ }\textbf {\bibinfo {volume} {421}},\ \bibinfo {pages}
  {611} (\bibinfo {year} {2003})}\BibitemShut {NoStop}%
\bibitem [{\citenamefont {Geddes}\ \emph {et~al.}(2008)\citenamefont {Geddes},
  \citenamefont {Nakamura}, \citenamefont {Plateau}, \citenamefont {Toth},
  \citenamefont {Cormier-Michel}, \citenamefont {Esarey}, \citenamefont
  {Schroeder}, \citenamefont {Cary},\ and\ \citenamefont {Leemans}}]{Geddes}%
  \BibitemOpen
  \bibfield  {author} {\bibinfo {author} {\bibfnamefont {C.~G.~R.}\
  \bibnamefont {Geddes}}, \bibinfo {author} {\bibfnamefont {K.}~\bibnamefont
  {Nakamura}}, \bibinfo {author} {\bibfnamefont {G.~R.}\ \bibnamefont
  {Plateau}}, \bibinfo {author} {\bibfnamefont {C.}~\bibnamefont {Toth}},
  \bibinfo {author} {\bibfnamefont {E.}~\bibnamefont {Cormier-Michel}},
  \bibinfo {author} {\bibfnamefont {E.}~\bibnamefont {Esarey}}, \bibinfo
  {author} {\bibfnamefont {C.~B.}\ \bibnamefont {Schroeder}}, \bibinfo {author}
  {\bibfnamefont {J.~R.}\ \bibnamefont {Cary}}, \ and\ \bibinfo {author}
  {\bibfnamefont {W.~P.}\ \bibnamefont {Leemans}},\ }\href {\doibase
  10.1103/PhysRevLett.100.215004} {\bibfield  {journal} {\bibinfo  {journal}
  {Phys. Rev. Lett.}\ }\textbf {\bibinfo {volume} {100}},\ \bibinfo {pages}
  {215004} (\bibinfo {year} {2008})}\BibitemShut {NoStop}%
\bibitem [{\citenamefont {Zhang}\ \emph {et~al.}(2015)\citenamefont {Zhang},
  \citenamefont {Khudik},\ and\ \citenamefont {Shvets}}]{zhangprl2015}%
  \BibitemOpen
  \bibfield  {author} {\bibinfo {author} {\bibfnamefont {X.}~\bibnamefont
  {Zhang}}, \bibinfo {author} {\bibfnamefont {V.~N.}\ \bibnamefont {Khudik}}, \
  and\ \bibinfo {author} {\bibfnamefont {G.}~\bibnamefont {Shvets}},\ }\href
  {\doibase http://dx.doi.org/10.1103/PhysRevLett.114.184801} {\bibfield
  {journal} {\bibinfo  {journal} {Phys. Rev. Lett.}\ }\textbf {\bibinfo
  {volume} {114}},\ \bibinfo {pages} {184801} (\bibinfo {year}
  {2015})}\BibitemShut {NoStop}%
\bibitem [{\citenamefont {Shaw}\ \emph {et~al.}(2017)\citenamefont {Shaw},
  \citenamefont {Lemos}, \citenamefont {Amorim}, \citenamefont
  {Vafaei-Najafabadi}, \citenamefont {Marsh}, \citenamefont {Tsung},
  \citenamefont {Mori},\ and\ \citenamefont {Joshi}}]{shawprl2017}%
  \BibitemOpen
  \bibfield  {author} {\bibinfo {author} {\bibfnamefont {J.~L.}\ \bibnamefont
  {Shaw}}, \bibinfo {author} {\bibfnamefont {N.}~\bibnamefont {Lemos}},
  \bibinfo {author} {\bibfnamefont {L.~D.}\ \bibnamefont {Amorim}}, \bibinfo
  {author} {\bibfnamefont {N.}~\bibnamefont {Vafaei-Najafabadi}}, \bibinfo
  {author} {\bibfnamefont {K.~A.}\ \bibnamefont {Marsh}}, \bibinfo {author}
  {\bibfnamefont {F.~S.}\ \bibnamefont {Tsung}}, \bibinfo {author}
  {\bibfnamefont {W.~B.}\ \bibnamefont {Mori}}, \ and\ \bibinfo {author}
  {\bibfnamefont {C.}~\bibnamefont {Joshi}},\ }\href {\doibase
  http://dx.doi.org/10.1103/PhysRevLett.118.064801} {\bibfield  {journal}
  {\bibinfo  {journal} {Phys. Rev. Lett.}\ }\textbf {\bibinfo {volume} {118}},\
  \bibinfo {pages} {064801} (\bibinfo {year} {2017})}\BibitemShut {NoStop}%
\bibitem [{\citenamefont {Mahieu}\ \emph {et~al.}(2018)\citenamefont {Mahieu},
  \citenamefont {Jourdain}, \citenamefont {Phuoc}, \citenamefont {Dorchies},
  \citenamefont {Goddet}, \citenamefont {Lifschitz}, \citenamefont {Renaudin},\
  and\ \citenamefont {Lecherbourg}}]{mahieu2018probing}%
  \BibitemOpen
  \bibfield  {author} {\bibinfo {author} {\bibfnamefont {B.}~\bibnamefont
  {Mahieu}}, \bibinfo {author} {\bibfnamefont {N.}~\bibnamefont {Jourdain}},
  \bibinfo {author} {\bibfnamefont {K.~T.}\ \bibnamefont {Phuoc}}, \bibinfo
  {author} {\bibfnamefont {F.}~\bibnamefont {Dorchies}}, \bibinfo {author}
  {\bibfnamefont {J.-P.}\ \bibnamefont {Goddet}}, \bibinfo {author}
  {\bibfnamefont {A.}~\bibnamefont {Lifschitz}}, \bibinfo {author}
  {\bibfnamefont {P.}~\bibnamefont {Renaudin}}, \ and\ \bibinfo {author}
  {\bibfnamefont {L.}~\bibnamefont {Lecherbourg}},\ }\href {\doibase
  https://doi.org/10.1038/s41467-018-05791-4} {\bibfield  {journal} {\bibinfo
  {journal} {Nature Communications}\ }\textbf {\bibinfo {volume} {9}},\
  \bibinfo {pages} {3276} (\bibinfo {year} {2018})}\BibitemShut {NoStop}%
\bibitem [{\citenamefont {Li}\ \emph {et~al.}(2018)\citenamefont {Li},
  \citenamefont {Wu},\ and\ \citenamefont {Wu}}]{li2018situ}%
  \BibitemOpen
  \bibfield  {author} {\bibinfo {author} {\bibfnamefont {L.}~\bibnamefont
  {Li}}, \bibinfo {author} {\bibfnamefont {Y.}~\bibnamefont {Wu}}, \ and\
  \bibinfo {author} {\bibfnamefont {J.}~\bibnamefont {Wu}},\ }\href {\doibase
  https://doi.org/10.1016/j.micron.2018.05.007} {\bibfield  {journal} {\bibinfo
   {journal} {Micron}\ }\textbf {\bibinfo {volume} {111}},\ \bibinfo {pages}
  {1} (\bibinfo {year} {2018})}\BibitemShut {NoStop}%
\bibitem [{\citenamefont {Wu}\ \emph {et~al.}(2016)\citenamefont {Wu},
  \citenamefont {Gao}, \citenamefont {Li}, \citenamefont {Parish},
  \citenamefont {Liu}, \citenamefont {Liaw},\ and\ \citenamefont
  {An}}]{wu2016intragranular}%
  \BibitemOpen
  \bibfield  {author} {\bibinfo {author} {\bibfnamefont {W.}~\bibnamefont
  {Wu}}, \bibinfo {author} {\bibfnamefont {Y.}~\bibnamefont {Gao}}, \bibinfo
  {author} {\bibfnamefont {N.}~\bibnamefont {Li}}, \bibinfo {author}
  {\bibfnamefont {C.~M.}\ \bibnamefont {Parish}}, \bibinfo {author}
  {\bibfnamefont {W.}~\bibnamefont {Liu}}, \bibinfo {author} {\bibfnamefont
  {P.~K.}\ \bibnamefont {Liaw}}, \ and\ \bibinfo {author} {\bibfnamefont
  {K.}~\bibnamefont {An}},\ }\href {\doibase
  https://doi.org/10.1016/j.actamat.2016.08.058} {\bibfield  {journal}
  {\bibinfo  {journal} {Acta Materialia}\ }\textbf {\bibinfo {volume} {121}},\
  \bibinfo {pages} {15} (\bibinfo {year} {2016})}\BibitemShut {NoStop}%
\bibitem [{\citenamefont {Mankowsky}\ \emph {et~al.}(2014)\citenamefont
  {Mankowsky}, \citenamefont {Subedi}, \citenamefont {F{\"o}rst}, \citenamefont
  {Mariager}, \citenamefont {Chollet}, \citenamefont {Lemke}, \citenamefont
  {Robinson}, \citenamefont {Glownia}, \citenamefont {Minitti}, \citenamefont
  {Frano} \emph {et~al.}}]{mankowsky2014nonlinear}%
  \BibitemOpen
  \bibfield  {author} {\bibinfo {author} {\bibfnamefont {R.}~\bibnamefont
  {Mankowsky}}, \bibinfo {author} {\bibfnamefont {A.}~\bibnamefont {Subedi}},
  \bibinfo {author} {\bibfnamefont {M.}~\bibnamefont {F{\"o}rst}}, \bibinfo
  {author} {\bibfnamefont {S.~O.}\ \bibnamefont {Mariager}}, \bibinfo {author}
  {\bibfnamefont {M.}~\bibnamefont {Chollet}}, \bibinfo {author} {\bibfnamefont
  {H.}~\bibnamefont {Lemke}}, \bibinfo {author} {\bibfnamefont {J.~S.}\
  \bibnamefont {Robinson}}, \bibinfo {author} {\bibfnamefont {J.~M.}\
  \bibnamefont {Glownia}}, \bibinfo {author} {\bibfnamefont {M.~P.}\
  \bibnamefont {Minitti}}, \bibinfo {author} {\bibfnamefont {A.}~\bibnamefont
  {Frano}},  \emph {et~al.},\ }\href {\doibase
  https://doi.org/10.1038/nature13875} {\bibfield  {journal} {\bibinfo
  {journal} {Nature}\ }\textbf {\bibinfo {volume} {516}},\ \bibinfo {pages}
  {71} (\bibinfo {year} {2014})}\BibitemShut {NoStop}%
\bibitem [{\citenamefont {Buzzi}\ \emph {et~al.}(2019)\citenamefont {Buzzi},
  \citenamefont {F{\"o}rst},\ and\ \citenamefont
  {Cavalleri}}]{buzzi2019measuring}%
  \BibitemOpen
  \bibfield  {author} {\bibinfo {author} {\bibfnamefont {M.}~\bibnamefont
  {Buzzi}}, \bibinfo {author} {\bibfnamefont {M.}~\bibnamefont {F{\"o}rst}}, \
  and\ \bibinfo {author} {\bibfnamefont {A.}~\bibnamefont {Cavalleri}},\ }\href
  {\doibase https://doi.org/10.1098/rsta.2017.0478} {\bibfield  {journal}
  {\bibinfo  {journal} {Philosophical Transactions of the Royal Society A}\
  }\textbf {\bibinfo {volume} {377}},\ \bibinfo {pages} {20170478} (\bibinfo
  {year} {2019})}\BibitemShut {NoStop}%
\bibitem [{\citenamefont {Arber}\ \emph {et~al.}(2015)\citenamefont {Arber},
  \citenamefont {Bennett}, \citenamefont {Brady}, \citenamefont
  {Lawrence-Douglas}, \citenamefont {Ramsay}, \citenamefont {Sircombe},
  \citenamefont {Gillies}, \citenamefont {Evans}, \citenamefont {Schmitz},
  \citenamefont {Bell} \emph {et~al.}}]{arber2015contemporary}%
  \BibitemOpen
  \bibfield  {author} {\bibinfo {author} {\bibfnamefont {T.}~\bibnamefont
  {Arber}}, \bibinfo {author} {\bibfnamefont {K.}~\bibnamefont {Bennett}},
  \bibinfo {author} {\bibfnamefont {C.}~\bibnamefont {Brady}}, \bibinfo
  {author} {\bibfnamefont {A.}~\bibnamefont {Lawrence-Douglas}}, \bibinfo
  {author} {\bibfnamefont {M.}~\bibnamefont {Ramsay}}, \bibinfo {author}
  {\bibfnamefont {N.}~\bibnamefont {Sircombe}}, \bibinfo {author}
  {\bibfnamefont {P.}~\bibnamefont {Gillies}}, \bibinfo {author} {\bibfnamefont
  {R.}~\bibnamefont {Evans}}, \bibinfo {author} {\bibfnamefont
  {H.}~\bibnamefont {Schmitz}}, \bibinfo {author} {\bibfnamefont
  {A.}~\bibnamefont {Bell}},  \emph {et~al.},\ }\href {\doibase
  http://dx.doi.org/10.1088/0741-3335/57/11/113001} {\bibfield  {journal}
  {\bibinfo  {journal} {Plasma Physics and Controlled Fusion}\ }\textbf
  {\bibinfo {volume} {57}},\ \bibinfo {pages} {113001} (\bibinfo {year}
  {2015})}\BibitemShut {NoStop}%
\bibitem [{\citenamefont {Horný}\ \emph {et~al.}(2017)\citenamefont {Horný},
  \citenamefont {Nejdl}, \citenamefont {Kozlová}, \citenamefont {Krůs},
  \citenamefont {Boháček}, \citenamefont {Petržílka},\ and\ \citenamefont
  {Klimo}}]{horny2017temporal}%
  \BibitemOpen
  \bibfield  {author} {\bibinfo {author} {\bibfnamefont {V.}~\bibnamefont
  {Horný}}, \bibinfo {author} {\bibfnamefont {J.}~\bibnamefont {Nejdl}},
  \bibinfo {author} {\bibfnamefont {M.}~\bibnamefont {Kozlová}}, \bibinfo
  {author} {\bibfnamefont {M.}~\bibnamefont {Krůs}}, \bibinfo {author}
  {\bibfnamefont {K.}~\bibnamefont {Boháček}}, \bibinfo {author}
  {\bibfnamefont {V.}~\bibnamefont {Petržílka}}, \ and\ \bibinfo {author}
  {\bibfnamefont {O.}~\bibnamefont {Klimo}},\ }\href {\doibase
  https://doi.org/10.1063/1.4985687} {\bibfield  {journal} {\bibinfo  {journal}
  {Physics of Plasmas}\ }\textbf {\bibinfo {volume} {24}},\ \bibinfo {pages}
  {063107} (\bibinfo {year} {2017})}\BibitemShut {NoStop}%
\end{thebibliography}%

\end{document}